# Two-photon Interface of Nuclear Spins Based on the Opto-Nuclear Quadrupolar Effect

Haowei Xu[1], Changhao Li[1,2], Guoqing Wang[1,2], Hua Wang[1], Hao Tang[3],
Ariel Rebekah Barr[3], Paola Cappellaro[1,2,4, †], and Ju Li[1,3, ‡]

[1] Department of Nuclear Science and Engineering, Massachusetts Institute of Technology, Cambridge, Massachusetts 02139, USA

[2] Research Laboratory of Electronics, Massachusetts Institute of Technology, Cambridge, MA 02139, USA

[3] Department of Materials Science and Engineering, Massachusetts Institute of Technology, Cambridge, Massachusetts 02139, USA

[4] Department of Physics, Massachusetts Institute of Technology, Cambridge, MA 02139, USA

* Corresponding authors: † pcappell@mit.edu, ‡ liju@mit.edu

## Abstract

Photons and nuclear spins are two well-known building blocks in quantum information science and technology. Establishing an efficient interface between optical photons and nuclear spins, while highly desirable for hybridizing these two quantum systems, is challenging because the interactions between nuclear spins and the environment are usually weak in magnitude, and there is also a formidable gap between nuclear spin frequencies and optical frequencies. In this work, we propose an opto-nuclear quadrupolar (ONQ) effect, whereby optical photons can be efficiently coupled to nuclear spins, similar to Raman scattering. Compared to previous works, ancilla electron spins are not required for the ONQ effect. This leads to advantages such as applicability in defect-free nonmagnetic crystals and longer nuclear spin coherence time. In addition, the frequency of the optical photons can be arbitrary, so they can be fine-tuned to minimize the material heating and to match telecom wavelengths for long-distance communications. Using perturbation theory and first-principles calculations, we demonstrate that the ONQ effect is stronger by several orders of magnitude than other nonlinear optical effects that could couple to nuclear spins. Based on this rationale, we propose promising applications of the ONQ effect, including quantum memory, quantum transduction, and materials isotope spectroscopy. We also discuss issues relevant to the experimental demonstration of the ONQ effect.

**Introduction.** In recent years, we have witnessed the rapid growth of quantum information science and technology, which may enable numerous applications with capabilities beyond their classical counterparts. These advances are enabled by the remarkable success of various qubit platforms, including superconducting circuits [1], trapped ions [2], semiconductor defects [3,4], and Rydberg atoms [5,6]. An ideal qubit system should simultaneously embody long coherence time, fast operation, and scalability. While nuclear spins have unparalleled coherence times [7,8], their control poses grand challenges due to weak interactions with the environment.

Traditionally, nuclear spins have been manipulated using nuclear magnetic resonance (NMR) techniques, whereby external magnetic fields couple to the nuclear magnetic moment which is linear in $I$, the nuclear spin angular momentum. Recently, alternative approaches for controlling nuclear spins at the microscopic/mesoscopic scale have been explored, using electron-nuclear spin interaction [9–14], microwave electric field [15–18], phonons/mechanical waves [19,20], etc. Of particular interest are optical approaches [21–25], which can be non-contact, ultra-strong, and ultrafast. Optical approaches are widely used not only in quantum technologies, including communication, sensing, and computing, but also in many other disciplines [26–29]. If an efficient interface between nuclear spins and optical photons can be established, then it would be possible to realize a hybrid platform for distributed quantum computing and long-distance quantum communications, combining long-coherence memory and strong interconnects that can significantly boost scalability.

Optical control over nuclear spins is hindered by the formidable gap between nuclear spin ($10^6$ ~ $10^9$ Hz) and optical frequencies (~ $10^{15}$ Hz). To address this, adjacent paramagnetic electron spins have been harnessed as the intermediary– electron spins interact with optical photons via (virtual) orbital transitions and then control nuclear spins via the hyperfine interaction. While this approach has been successfully applied [21–25], e.g., in color centers in diamond and rare-earth-doped semiconductors, it suffers from several limitations. First, it requires the existence of nearby localized electron spins with $S \neq 0$, which only holds for defect systems or magnetic materials. Besides, the coherence time of electron spin is usually much shorter than that of nuclear spin, thus fast operation is required. The existence of unpaired electron spins also shortens the nuclear spin coherence time [11,30]. The strength of the hyperfine interaction decays with distance, so an electron spin can only control a relatively small nuclear spin ensemble. Finally, the interface between electron spins and optical photons is sensitive to the environment (material inhomogeneities, electron-phonon coupling, etc.), substantially limiting the fidelity of the entanglement between remote spins. To this end, it is highly desirable to introduce other optical mechanisms that can couple to nuclear spins without directly involving localized electron spins.

In this work, we overcome these difficulties by proposing an opto-nuclear quadrupolar (ONQ) effect, which is second-order in the *electric* field and nuclear spin $I$, as mediated by the quadrupole electric coupling, and is thus one of the nonlinear optical (NLO) responses of materials [31,32] present in perfect crystals. Via the ONQ effect, nuclear spins can be coherently controlled by two-color photons, without electron spins as the media. The basic mechanism of such control is that optical photons affect the total electron density distribution and the electrical field gradient (EFG) at the site of the nuclear spin. By applying two-color photons with respective frequencies $\omega_{o1}$ and $\omega_{o2}$, the nuclear quadrupole interaction, which is linear in EFG, would oscillate at the frequency $|\omega_{o1} - \omega_{o2}|$, which can be tuned to match the nuclear spin resonance frequency and trigger nuclear spin transitions. We remark that (1) ancilla electron spins with $S \neq 0$ are not necessary for the ONQ effect, therefore the aforementioned limitations due to the electron spins can be eased; (2) the optical frequencies $\omega_{o1}$ and $\omega_{o2}$ can be arbitrary, so they can be tuned to minimize material heating effects and to match telecom wavelength, fulfilling the potential of long-distance information transmission required in various quantum information applications. In terms of formalism, the ONQ effect is similar to Raman scattering, which is also a two-photon process whereby optical photons couple to the electronic orbitals and consequently losing/gaining energy to a phonon mode in crystal or rotational/vibrational mode in molecules, whose typical energies (~$10^{12}$ Hz) are much lower than the optical frequencies (~$10^{15}$ Hz). In ONQ, the energy exchange is to nuclear-spin mode instead of phonons. Indeed, both nuclear spin and phonon couple electrically to the electronic orbitals, the former through the nuclear electric quadrupole moment and EFG, and the latter through the Born effective charge. As we will show, the ONQ transition rate can be as high as $10^{-3} \times$ as Raman scattering for the same light intensity.

In the following, we first introduce the mechanism of the ONQ effect. Using both perturbation theory and ab initio density functional theory (DFT) [33,34] calculations, we estimate the strength of the ONQ response and show that it is much stronger than other NLO effects that could couple optical photons and nuclear spins, such as the nonlinear nuclear Zeeman interaction (Table I). In this regard, we suggest several promising applications of the ONQ effect under feasible experimental conditions (Table II). First, we demonstrate that the ONQ effect can drive Rabi oscillations of a single nuclear spin. Then, we will show the potential of ONQ for material spectroscopy and isotope mapping, as two-photon microscopy [35] is a well-established technique. Then, leveraging the optical interface established by the ONQ effect, we show the quantum information carried by optical photons can be directly stored in a nuclear spin ensemble (NSE) quantum memory, where a large number ($\gtrsim 10^{10}$) of nuclear spins can be collectively excited. Since the nuclear spins can also be coupled to microwave (MW)/radiofrequency (RF) photons through the Zeeman interaction, they can serve as the media for quantum transduction between MW/RF and optical photons, which is of critical importance for hybrid quantum systems [36]. By solving the master equation, we show that a transduction fidelity of over 90% can be achieved, benefiting from the long coherence time of nuclear spins. We will also discuss some issues relevant to the experimental demonstration of the ONQ effect, including proof-of-principles experiments, possible approaches to readout the nuclear spin states, and possible challenges in demonstrating the ONQ effect.

**Opto-nuclear quadrupolar effect**. The Hamiltonian of a nuclear spin includes both a magnetic (Zeeman) interaction term $H_Z$ and an electric quadrupole interaction term $H_Q$, expressed as

$$H = H_Z + H_Q = g_m \sum_i \mathcal{B}_i I_i + \sum_{ij} Q_{ij} I_i I_j \qquad (1)$$

where $i, j = x, y,$ or $z$ are Cartesian indices, $I_i$ is the nuclear angular momentum operator, $\mathcal{B}$ is the local magnetic field, and $g_m \sim 10 \times \frac{2\pi \cdot \text{MHz}}{\text{T}}$ is the gyromagnetic ratio of the nucleus under consideration. The quadrupole tensor $Q_{ij}$ can be expressed as

$$Q_{ij} = \frac{eq\mathcal{V}_{ij}}{2I(2I-1)} \qquad (2)$$

where $e$ is the electron charge, $q$ is the nuclear electric quadrupole moment, and $I$ is the nuclear angular momentum quantum number. $Q_{ij}$ is nonzero only when $I > \frac{1}{2}$. $\mathcal{V}$ is the EFG tensor at the nuclear site. The quadrupole interaction energy scale is determined by the factor $C_q \equiv \frac{eq\mathcal{V}_{zz}}{2\pi}$, (Planck constant $\hbar = 1$), where $\mathcal{V}_{zz}$ is the largest principal value of the $\mathcal{V}$ tensor. $C_q$ ranges between tens of kHz to GHz. For example, for $^{69}$Ga or $^{71}$Ga nuclei in wurtzite GaN, $C_q \sim 1$ MHz; for $^{177}$Hf or $^{179}$Hf nuclei in HfO$_2$, $C_q \sim 1$ GHz; for $^{181}$Ta defects in hexagonal BN, $C_q$ can reach 7 GHz, according to our ab initio calculations. Generally, $C_q$ is larger in magnitude for isotopes with large quadrupole moments residing in highly local asymmetric environments. It is also possible to improve the $C_q$ by strain or (point) defects, which could enhance the structural asymmetry (Appendix A.5). Note that in Eq. (1), we do not include terms dependent on the electron spin $S$, such as the hyperfine interaction, thus Eq. (1) applies to systems without unpaired electron spins.

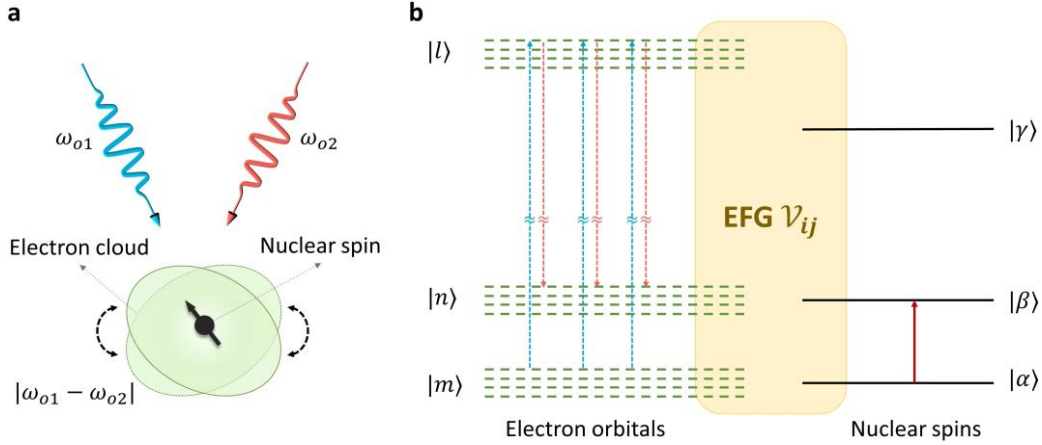

**Figure 1** Illustration of the ONQ effect. **(a)** A semi-classical illustration of the ONQ effect. Under two photons with frequencies $\omega_{o1}$ and $\omega_{o2}$, respectively, the electron cloud vibrates with frequency $|\omega_{o1} - \omega_{o2}|$ and modulates the quadrupolar interaction of the nuclear spin. **(b)** Quantum energy level diagram of the ONQ effect. Electrons do (virtual) transitions between three orbitals and modulate the EFG at the nuclear site. Nuclear spins can transit between two energy levels if the frequency matching condition is satisfied.

Traditional approaches to manipulate nuclear spins such as NMR rely on the magnetic interaction $H_Z$, which is linear in $\mathcal{B}$ – static DC fields yield the Zeeman splitting and Larmor precession, and resonant AC fields induce Rabi oscillation. However, even in the presence of large magnetic fields, the nuclear spin frequencies can only achieve a few hundred MHz to GHz, much smaller than the optical frequencies. Therefore, an optical photon cannot directly drive resonant nuclear spin transitions, although it has an associated magnetic field. Still, resonance can be achieved when simultaneously applying two-color lasers whose frequency difference matches nuclear spin frequency. This can be considered the nonlinear nuclear Zeeman (NNZ) effect. From perturbation theory, one can obtain the NNZ interaction strength as $\mathcal{G}^{\mathrm{NNZ}} \sim \frac{g_m^2}{\Delta}$, where $\Delta$ is the detuning from resonant frequency. Since the optical frequency $\omega_o$ is much higher than the nuclear spin frequency, one has $\Delta \approx \omega_o \sim 1$ eV. This leads to $\mathcal{G}^{\mathrm{NNZ}} \sim 10^{-4} \times \frac{2\pi \cdot \mathrm{MHz}}{(\mathrm{V/Å})^2}$, which is too weak for any practical applications. Here we have converted the magnetic field strength $\mathcal{B}$ of an optical photon to its electric field strength $\mathcal{E}$ using $\mathcal{E} = c_0 \mathcal{B}$, where $c_0$ is the speed of light.

In contrast, the relatively less-explored nuclear electric quadrupole interaction $H_Q$ does not depend on the magnetic field, but instead on the electric field, which could enable more efficient coupling between optical photons and nuclear spins. The EFG originates in the electric potential generated by surrounding electrons [37] and is a functional of the electron density $\rho$. If an external electric field $\mathcal{E}$ is applied, then electrons will redistribute in real space (a semi-classical illustration of the electron distribution is depicted in Figure 1a), and $\rho$ can be perturbatively expressed as $\rho = \sum_{\alpha=0} \rho^{(\alpha)}$, where $\rho^{(0)}$ is the equilibrium density, while $\rho^{(\alpha)} \propto \mathcal{E}^\alpha$ is the $\alpha$-th order perturbation. Since the EFG, and hence the quadrupole tensor, is a functional of $\rho$, one has $Q_{ij} = Q_{ij}^{(0)} + \mathcal{C}_{ij}^p \mathcal{E}_p + \mathcal{D}_{ij}^{pq} \mathcal{E}_p \mathcal{E}_q + \cdots$. Here $p$ and $q$ label the polarization of the electric fields, while $\mathcal{C}_{ij}^p \equiv \frac{\partial Q_{ij}}{\partial \mathcal{E}_p}$ and $\mathcal{D}_{ij}^{pq} \equiv \frac{\partial^2 Q_{ij}}{\partial \mathcal{E}_p \partial \mathcal{E}_q}$ are the first- and second-order response functions, respectively. The first-order response $\mathcal{C}_{ij}^p \mathcal{E}_p$ leads to the so-called nuclear electric resonance (NER) [16–18], whereby an MW/RF electric field drives the Rabi oscillation of nuclear spins. Experimental results and our theoretical predictions indicate that $\mathcal{C}_{ij}^p \sim 10 \frac{2\pi \cdot \text{MHz}}{\text{V/Å}}$ (Appendix B.2.2). Interestingly, the second-order perturbation theory effect from $\mathcal{C}_{ij}^p \mathcal{E}_p$, which is $\mathcal{G}^{\text{NNER}} \sim \frac{\mathcal{C}_{ij}^p \mathcal{C}_{lm}^q \mathcal{E}_p \mathcal{E}_q}{\Delta}$, corresponds to the nonlinear coupling between nuclear spins and two-color photons. This can be considered as the nonlinear nuclear electric resonance (NNER). However, the interaction strength of NNER is only $\mathcal{G}^{\text{NNER}} \sim 10^{-6} \times \frac{2\pi \cdot \text{MHz}}{(\text{V/Å})^2}$, even weaker than the NNZ coupling.

The second-order response in $Q$, which is $\mathcal{D}_{ij}^{pq} \mathcal{E}_p \mathcal{E}_q$, could establish a more efficient interface between optical photons and nuclear spins. Under two electric fields $\mathcal{E}(\omega_{o1})e^{i\omega_{o1}t}$ and $\mathcal{E}(-\omega_{o2})e^{-i\omega_{o2}t}$, the quadrupole interaction $H_Q$ becomes

$$\begin{aligned} H_Q(t) &= H_Q^{(0)} + H_Q^{(2)}(t; \omega_{o1}, -\omega_{o2}) \\ &= \sum_{ij} Q_{ij}^{(0)} I_i I_j + \mathcal{D}_{ij}^{pq}(\omega_{o1} - \omega_{o2}; \omega_{o1}, -\omega_{o2}) \mathcal{E}_p(\omega_{o1}) \mathcal{E}_q(-\omega_{o2}) I_i I_j e^{i(\omega_{o1} - \omega_{o2})t} + h.c. \end{aligned} \quad (3)$$

Here $h.c.$ stands for Hermitian conjugate. We have omitted terms with frequencies $\omega_{o1}$, $\omega_{o2}$ and $\omega_{o1} + \omega_{o2}$, which are rapidly oscillating in comparison to the nuclear spin energy and are thus decoupled from nuclear spin dynamics in the spirit of the rotating wave approximation.

$Q_{ij}^{(0)}$ is the intrinsic quadrupole tensor at $\mathcal{E}=0$. One can see that $H_Q^{(2)}(t;\omega_{o1},-\omega_{o2})$ oscillates at frequency $\omega_{o1}-\omega_{o2}$, which can be tuned to match the nuclear spin energy scale. This is the ONQ effect.

The mechanism of the ONQ effect is as follows. The total electron density and its long-range Coulombic coupling with the EFG at the nuclear site serve as the bridge between optical photons and nuclear spins. Under light illumination, electrons undergo (virtual) transitions between three orbitals [Figure 1b, see also Eq. (4) below] and modulate the EFG at the nuclear site. Since the energy scales involved with electron orbital energy levels are typically on the order of 1 eV, (near-)resonance can be achieved. Notably, unpaired electronic spin $S$ and magnetic interaction in general are not necessary for the quadrupole interaction, and the limitations from ancilla electrons spins mentioned at the beginning can be eased.

**Table I.** Orders of interactions for various NLO effects that could couple optical photons and nuclear spins. Interactions involving nuclear spins are marked in red.

|  | **Electron Electric Dipole Moment** | **Nuclear Magnetic Dipole Moment** | **Nuclear Electric Quadrupole Moment** | Strength $\left[\frac{2\pi\cdot\text{MHz}}{(\text{V/Å})^2}\right]$ |
|---|---|---|---|---|
| **ONQ** | Second | / | First | $10\sim 10^2$ |
| **NNZ** | / | Second | / | $\sim 10^{-4}$ |
| **NNER** | Second | / | Second | $\sim 10^{-6}$ |
| Strength | $\sim 10^8 \times \frac{2\pi\cdot\text{MHz}}{\text{V/Å}}$ (a) | $\sim 10^2 \times \frac{2\pi\cdot\text{MHz}}{\text{V/Å}}$ (b) | $[1\sim 10^3]\times 2\pi\cdot\text{MHz}$ | / |

(a) The electric dipole of electrons is typically on the order of $1e\cdot\text{Å}$, and 1 eV is equivalent to $2.4\times 10^8 \times 2\pi\cdot\text{MHz}$.

(b) Magnetic field strength $\mathcal{B}$ is converted to electric field strength $\mathcal{E}$ using $\mathcal{B}=\frac{\mathcal{E}}{c_0}$, where $c_0$ is the speed of light.

Remarkably, the ONQ effect can be much stronger than the NNZ or NNER effects. In the next section, we will use both perturbation theory and ab initio calculations to estimate the ONQ interaction strength, and show that it can reach $10^2 \times \frac{2\pi\cdot\text{MHz}}{(\text{V/Å})^2}$, greater than $\mathcal{G}^{\text{NNZ}}$ or $\mathcal{G}^{\text{NNER}}$ by several orders of magnitude. To understand this phenomenon, we need to distinguish three physical interactions, namely (1) electron orbital interaction between external electric fields and the electron electric dipole moment, (2) nuclear magnetic (Zeeman) interaction between external magnetic fields and the nuclear magnetic dipole moment, and (3) nuclear quadrupole interaction between electrons and the nuclear electric quadrupole moment. Both the nuclear Zeeman and the nuclear quadrupole interactions are $I$-dependent. The strengths of these interactions are listed at the bottom line of Table I. Notably, the electric field interaction with electron electric dipole moment tend to be stronger than the interactions involving nuclear magnetic/electric moments. ONQ, NNZ, and NNER involve different orders of these factors, which are shown in Table I. One can see that ONQ is second-order in the electron electric dipole moment, but first-order in the nuclear electric quadrupole moment. In contrast, both NNZ and NNER are second-order in the interaction involving nuclear magnetic dipole or electric quadrupole moments, which makes them significantly weaker than the ONQ effect.

**Estimation of the $\mathcal{D}$ tensor**. In this section, we estimate the order of magnitude of $\mathcal{D}$, which determines the strength of the ONQ effect. We use wurtzite gallium nitride (wGaN) as an example, while the estimation of the $\mathcal{D}$ tensor of some other compounds can be found in Appendix B. wGaN has a wide bandgap of ~3.4 eV and is advantageous for opto-electronic and opto-nuclear applications [38]. We focus on the Ga nuclei. In the following, we will use both second-order perturbation theory and DFT calculations to predict the magnitude of the $\mathcal{D}$ tensor. An assessment of the validity of these theoretical predictions can be found in the Appendix B.2, which also contains the theoretical prediction on the strength of the NER (the $\mathcal{C}$ tensor).

In the single-particle approximation, one can obtain the $\mathcal{D}$ tensor from second-order perturbation theory as [31]

$$\mathcal{D}_{ij}^{pq}(\omega_{o1} - \omega_{o1}; \omega_{o1}, -\omega_{o2})$$
$$= \frac{e^3 q}{2I(2I-1)} \sum_{mnl} \frac{[\mathcal{V}_{ij}]_{mn}}{E_{mn} - \hbar(\omega_{o1} - \omega_{o2}) + i\eta} \times \left\{ \frac{f_{lm}[r_p]_{nl}[r_q]_{lm}}{E_{ml} - \hbar\omega_{o2} + i\eta} - \frac{f_{nl}[r_q]_{nl}[r_p]_{lm}}{E_{ln} - \hbar\omega_{o2} + i\eta} \right\} + (p, o1 \leftrightarrow q, o2),\quad(4)$$

where $(p, o1 \leftrightarrow q, o2)$ indicates the simultaneous exchange of the $(p, o1)$ and $(q, o2)$ subscripts, which symmetrizes the two optical fields. $m, n, l$ are labels of the electronic states, $E_{mn} \equiv E_m - E_n$ and $f_{mn} = f_m - f_n$ are the energy and occupation differences between two electronic states $|m\rangle$ and $|n\rangle$, respectively. $[r_i]_{mn} \equiv \langle m|r_i|n\rangle$ is the position operator of the electrons. Meanwhile, $[\mathcal{V}_{ij}]_{mn} = \frac{e}{4\pi\varepsilon_0} \left\langle m \left| \frac{3r_i r_j - \delta_{ij} r^2}{r^5} \right| n \right\rangle$ is the EFG operator of the electrons in the single-particle approximation, where $\varepsilon_0$ is the vacuum permittivity. $\eta$ is the electron linewidth, which is on the order of 1 meV [39–41]. One can see that electrons do (virtual) three-band transitions under the two-color laser (Figure 1b). When $\omega_{o1\,(o2)} < E_g$ with $E_g$ the bandgap, resonant electron interband transitions cannot happen, and the $(m, n, l)$ pair that satisfies $E_{mn} = E_{ml} = E_g$ would make the major contribution to the $\mathcal{D}$ tensor. For an order-of-magnitude estimation of $\mathcal{D}$, we only consider this $(m, n, l)$ pair. We also use $[r_i]_{mn} \approx a_0$ and $\left\langle m \left| \frac{3r_i r_j - \delta_{ij} r^2}{r^5} \right| n \right\rangle \approx \frac{1}{a_0^3}$ in Eq. (4). Here $a_0$ is the Bohr radius, which is also approximately half the bond length in typical compounds and characterizes the spatial extent of the electron wavefunction in molecule or solid-state systems. We also ignore $\omega_{o1} - \omega_{o2}$ and $\eta$ in the denominator, since they are much smaller than $E_{mn}$, which is above 1 eV in typical semiconductors. Then, one has $\mathcal{D} \sim \frac{g_S}{2I(2I-1)} \frac{e^4 q}{4\pi\varepsilon_0 a_0} \frac{1}{E_g(E_g - \omega_{o1})}$, where $g_S = 2$ is the electron spin degeneracy. Using this relationship, we obtain $\mathcal{D} \sim 6 \times \frac{2\pi \cdot \text{MHz}}{(\text{V/Å})^2}$ for the Ga nuclei in wGaN when $E_g - \omega_{o2} = 0.2$ eV. Similar to other nonlinear optical effects such as second-harmonics generation, when $\omega_{o1\,(o2)}$ is larger than the electronic bandgap $E_g$, electrons can undergo resonant transitions, which would significantly boost the transition rate and hence the response function $\mathcal{D}$. However, this would also lead to strong absorption of the laser energy and significant heating effect, which could damage the sample. Hence, we only consider $\omega_{o1\,(o2)} < E_g$ in the following.

Next, we will use DFT calculations to estimate the magnitude of $\mathcal{D}$, which give more detailed information on the quadrupole interaction. Some details on the DFT calculations can be found in Appendix B. Due to the P6$_3$mc symmetry (Figures 2a, 2b), the only nonzero components of the EFG tensor $\mathcal{V}_{ij}$ of Ga nuclei are $\mathcal{V}_{zz} = -2\mathcal{V}_{xx} = -2\mathcal{V}_{yy} \approx 3.34$ V/Å$^2$ in our DFT calculations, with $z$ along the $c$-axis of the wurtzite structure. Then we apply a homogeneous finite electric field $\mathcal{E}$ to calculate the change in EFG tensor $\Delta \mathcal{V}_{ij}$. In Figure 2c we plot $\Delta \mathcal{V}_{ij}$ as a function of $\mathcal{E}_x$. Notably, the first-order response $\left. \frac{\partial \mathcal{V}_{ij}}{\partial \mathcal{E}_x} \right|_{\mathcal{E}_x = 0}$ is zero for certain elements such as $\mathcal{V}_{xx}$ (inset of Figure 2c). This is because wGaN has mirror symmetry $\mathcal{M}_x$, leading to $\left. \frac{\partial \mathcal{V}_{xx}}{\partial \mathcal{E}_x} \right|_{\mathcal{E}_x = 0} = -\left. \frac{\partial \mathcal{V}_{xx}}{\partial \mathcal{E}_x} \right|_{\mathcal{E}_x = 0}$ and $\left. \frac{\partial \mathcal{V}_{xx}}{\partial \mathcal{E}_x} \right|_{\mathcal{E}_x = 0} = 0$. Note that the first-order response can be entirely forbidden if the system has inversion symmetry and the nucleus under consideration is located at the inversion center. The second-order responses, on the other hand, are not constrained by mirror symmetries. By fitting the raw data (solid points) with second-order polynomials (solid curves), we find $\left. \frac{\partial^2 \mathcal{V}_{ij}}{\partial \mathcal{E}_x^2} \right|_{\mathcal{E}_x = 0} \sim 10 \times \frac{\text{V/Å}^2}{(\text{V/Å})^2}$ and $\mathcal{D} = \left. \frac{\partial^2 Q_{ij}}{\partial \mathcal{E}_x^2} \right|_{\mathcal{E}_x = 0} = \frac{eq}{2I(2I-1)} \left. \frac{\partial^2 \mathcal{V}_{ij}}{\partial \mathcal{E}_x^2} \right|_{\mathcal{E}_x = 0} \sim 1 \times$

$\frac{2\pi \cdot \text{MHz}}{(\text{V/Å})^2}$ for Ga nuclei in wGaN, which is of the same order of magnitude as the estimation above using perturbation theory. Here we would like to remark that the electric field calculations in DFT are usually not very accurate because they involve the excited states of electrons (i.e., conduction bands), which DFT cannot describe accurately, as DFT is a groundstate theory [33]. Hence, the DFT calculation here should be considered as an order-of-magnitude estimation of the $\mathcal{D}$ tensor.

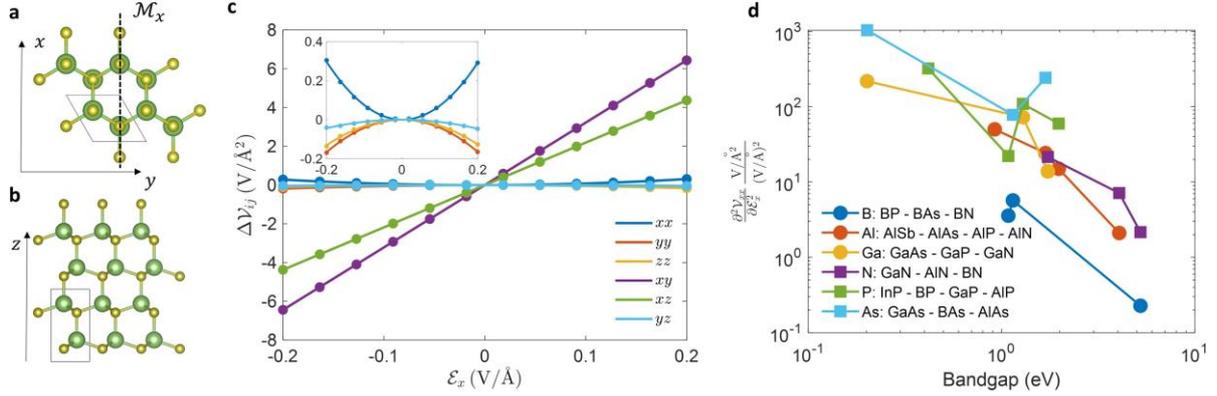

**Figure 2 (a, b)** Atomic structure of wurtzite GaN. Dashed line in (a) labels the mirror symmetry $\mathcal{M}_x$ of wGaN. Green: Ga; Yellow: N. **(c)** change in EFG tensor $\Delta \mathcal{V}_{ij}$ of Ga nuclei as a function of $\mathcal{E}_x$. Due to the $\mathcal{M}_x$ symmetry, four of six components of the $\mathcal{V}_{ij}$ tensor have first-order response $\frac{\partial \mathcal{V}_{ij}}{\partial \mathcal{E}_x}\big|_{\mathcal{E}_x=0} = 0$ (inset). **(d)** $\frac{\partial^2 \mathcal{V}_{xx}}{\partial \mathcal{E}_x^2}\big|_{\mathcal{E}_x=0}$ against the bandgap $E_g$ for all wurtzite III-V materials with $E_g > 0$ in DFT calculations.

Besides, the $\frac{\partial^2 \mathcal{V}}{\partial \mathcal{E}^2}$ tensor calculated using DFT is the static response ($\omega_{o1} = \omega_{o2} = 0$). Indeed, the evaluation of the full tensor $\mathcal{D}_{ij}^{pq}(\omega_{o1} - \omega_{o2}; \omega_{o1}, -\omega_{o2})$ at arbitrary frequencies using DFT calculations is not straightforward and is left for a future study. Still, the static response $\mathcal{D}(0;0,0)$ should be a lower bound for its value when $\omega_{o1}/\omega_{o2}$ is in the optical frequencies. This is because electron responses can be faster when $\omega_{o1}/\omega_{o2}$ are closer to the bandgap $E_g$, which is also manifested in Eq. (4). To further verify this point in DFT calculations, we use materials with narrower bandgaps. Intuitively, this would enhance the static responses because $\omega_{o1} = \omega_{o2} = 0$ is closer to $E_g$. In Figure 2d we plot the calculated $\frac{\partial^2 \mathcal{V}_{xx}}{\partial \mathcal{E}_x^2}$ against the bandgap $E_g$ for all wurtzite III-V materials with $E_g > 0$ in DFT calculations (DFT calculations usually underestimate bandgaps). One can observe a clear trend, where $\frac{\partial^2 \mathcal{V}_{xx}}{\partial \mathcal{E}_x^2}$ is bigger when $E_g$ is smaller. For GaAs with $E_g \sim 0.2$ eV in our calculation, $\frac{\partial^2 \mathcal{V}_{xx}}{\partial \mathcal{E}_x^2}$ can reach $10^3 \times \frac{\text{V/Å}^2}{(\text{V/Å})^2}$.

Besides $\frac{\partial^2 \mathcal{V}}{\partial \mathcal{E}^2}$, the $\mathcal{D} \equiv \frac{\partial^2 Q}{\partial \mathcal{E}^2}$ tensor also depends on the quadrupole moment $q$ and angular momentum $I$ of the nucleus under consideration. For nuclei with small $q$ such as Ga, one has $\mathcal{D} \sim [1 \sim 10^2] \times \frac{2\pi \cdot \text{MHz}}{(\text{V/Å})^2}$. For nuclei with large $q$ such as Ta and Hf, $\mathcal{D}$ can reach $[10 \sim 10^3] \times \frac{2\pi \cdot \text{MHz}}{(\text{V/Å})^2}$ or even larger values. Hence, for single nuclear spin manipulation, one could choose defect atoms with large $q$ such as Hf, which would give a large $\mathcal{D} \sim 10^2 \times \frac{2\pi \cdot \text{MHz}}{(\text{V/Å})^2}$. When instead looking for nuclear spin ensembles, one needs abundant isotopes in typical semiconductors, such as $^{75}$As in GaAs, whereby the $\mathcal{D}$ tensor could be smaller. For clarity, we will use $\mathcal{D} = 20 \times \frac{2\pi \cdot \text{MHz}}{(\text{V/Å})^2}$ for the discussion hereafter, which is suitable for $^{75}$As in GaAs, although larger $\mathcal{D}$ can be achievable in practice.

In the following sections, we will introduce several promising applications of the ONQ effect, including spectroscopy, quantum memory, and quantum transduction. A summary of these applications is shown in Table II.

**Table II.** List of potential applications of the opto-nuclear quadrupole effect. The desired transition(s) for a certain application are marked in red.

| Application | Specification | $\omega_{o1}$-photon (Optical) | $\omega_{o2}$-photon (Optical) | $\omega_m$-photon (MW/RF) | Nuclear Spin | Frequency Matching Condition |
|---|---|---|---|---|---|---|
| Single Nuclear Spin Manipulation | Rabi Oscillation | Pump | Pump | Not Involved | $\|e\rangle \leftrightarrow \|g\rangle$ | $\|\omega_{o1} - \omega_{o2}\| = \Delta_{ge}$ |
| Spectroscopy | Isotope Mapping, etc. | Pump | Creation | Not Involved | $\|e\rangle \leftrightarrow \|g\rangle$ | $\|\omega_{o1} - \omega_{o2}\| = \Delta_{ge}(X)$ |
| Quantum Memory | Storage | Annihilation | Pump | Not Involved | $\|G\rangle \to \|E\rangle$ | $\|\omega_{o1} - \omega_{o2}\| = \Delta_{GE}$ |
| Quantum Memory | Readout | Creation | Pump | Not Involved | $\|E\rangle \to \|G\rangle$ | $\|\omega_{o1} - \omega_{o2}\| = \Delta_{GE}$ |
| Quantum Transduction | Optical → MW/RF | Pump | Annihilation | Creation | Off-resonance | $\omega_{o1} - \omega_{o2} = \omega_m$, $\omega_{o1} - \omega_{o2} \neq \Delta_{GE}$ |
| Quantum Transduction | MW/RF → Optical | Pump | Creation | Annihilation | Off-resonance | $\omega_{o1} - \omega_{o2} = \omega_m$, $\omega_{o1} - \omega_{o2} \neq \Delta_{GE}$ |

**Single nuclear spin manipulation.** The ONQ Hamiltonian $H_Q^{(2)}(t; \omega_{o1}, -\omega_{o2})$, which oscillates in time with frequency $|\omega_{o1} - \omega_{o2}|$, enables the coupling between nuclear spins and optical photons. To better illustrate the basic mechanisms of the ONQ effect, we first demonstrate how it can drive the Rabi oscillation of a single nuclear spin. While nuclei with $I > \frac{1}{2}$ can have multiple nuclear spin energy levels, we focus on two nuclear spin eigenstates $|g\rangle$ and $|e\rangle$ with $H_Q^{(0)}|g\rangle = \epsilon_g|g\rangle$ and $H_Q^{(0)}|e\rangle = \epsilon_e|e\rangle$, and other nuclear spin states can be ignored since they are off-resonance. Here we apply the $\omega_{o1}$- and $\omega_{o2}$-laser simultaneously with frequency matching condition $|\omega_{o1} - \omega_{o2}| = \Delta_{ge}$, where $\Delta_{ge} \equiv \epsilon_e - \epsilon_g$ is the nuclear spin splitting. The frequency matching condition indicates that the excess energy of the two photons $|\omega_{o1} - \omega_{o2}|$ is absorbed (emitted) by the nuclear spins during the $|g\rangle \to |e\rangle$ ($|e\rangle \to |g\rangle$) transitions. This is similar to other NLO processes such as difference frequency generation or Raman scattering, where the excess energy is absorbed (emitted) by a third photon or a phonon. In the rotating frame, the ONQ Hamiltonian is $H_o = g_o \mathcal{E}(\omega_{o1})\mathcal{E}(-\omega_{o2})|g\rangle\langle e| + h.c.$, where $g_o \equiv \sum_{ij} \mathcal{D}_{ij} \langle g|I_i I_j|e\rangle$. The Rabi frequency between $|g\rangle$ and $|e\rangle$ is then $f_{g \leftrightarrow e} = g_o \mathcal{E}(\omega_{o1})\mathcal{E}(-\omega_{o2})$. With $\mathcal{E}(\omega_{o1}) = \mathcal{E}(-\omega_{o2}) = 0.5$ MV/cm, one has $f_{g \leftrightarrow e} \sim 100$ Hz. Considering that the decoherence rate of the nuclear spin can be lower than 1 Hz even at room temperature [7,8,42], an electric field of 0.5 MV/cm should suffice to demonstrate the manipulation of single nuclear spin before it gets decoherent. The only requirements for realizing such a Rabi oscillation are (1) nuclear spin with non-zero $\mathcal{D}$ tensor and (2) two-color laser satisfying the frequency matching condition, so it can serve as a proof-of-principle experiment for the ONQ effect. To detect the quantum state of the single nuclear spin, one possible approach is to introduce an electron spin for readout [11,16,43] after the Rabi oscillation induced by the ONQ interaction is finished. This can be realized by ionizing/neutralizing the defect atom that hosts the single nuclear spin, which has been demonstrated in e.g., Refs. [11,16]. Note that the single-electron transistor used in Refs. [11,16] required metallic structures with hundreds of nanometer size, which could influence the light propagation. While some specialized metallic structures have been used to enhance properties of light propagation and intensity, this might not be compatible with the design needed for SET and adverse effects might arise. An alternative readout strategy would be to work with systems that naturally have isolated electron spins. For example, the nitrogen nuclear spin in nitrogen-vacancy (NV) centers in diamond can be read out by detecting the transition energies of NV electron spins, which depend on the nuclear spin states due to their hyperfine interaction. Note that the frequency of the laser used for the ONQ interaction should be kept below the electronic transition energy (637 nm for NV⁻) to avoid unwanted absorption and heating effects. In the long term, it might be desirable to get rid of electron spins even in such cases. The possibility of an all-optical control over single nuclear spin is discussed in Appendix A.2.1.

**Table III.** Comparison between ONQ and Raman spectroscopy.

|  | Coupling strength (per formula unit) | Frequency shift | Linewidth |
|---|---|---|---|
| **ONQ** | $\sim 10^2 \times \frac{2\pi \cdot \text{MHz}}{(\text{V/Å})^2}$ | MHz - GHz | kHz |
| **Raman** | $\sim 10^5 \times \frac{2\pi \cdot \text{MHz}}{(\text{V/Å})^2}$ | THz | sub-THz |

**Spectroscopy.** The ONQ effect can potentially be used for material spectroscopy and isotopic mapping applications, similar to the well-established two-photon microscopy [35]. The scheme is as follows. Under a $\omega_{o1}$-laser, nuclear spins can undergo a spin flip provided the energy mismatch is released in the form of an emitted photon with shifted frequency $\omega_{o2}$. $\omega_{o2}$ needs to satisfy the frequency matching condition $|\omega_{o1} - \omega_{o2}| = \Delta_{ge}(X)$, where $\Delta_{ge}(X)$ is the nuclear spin resonance energy of isotope $X$ in a given material. The shifted side peaks at $\omega_{o2}$ can be the fingerprint of the isotopes – if a certain isotope $X$ is present, then one can detect two side peaks with frequencies $\omega_{o2} = \omega_{o1} \pm \Delta_{ge}(X)$. Note that $\Delta_{ge}(X)$ is also dependent on the chemical environment of isotope $X$. This is similar to the material characterization using Raman scattering, whereby a certain vibrational mode can be detected by measuring the position of the (anti-)Stokes peaks. We would like to remark that such ONQ spectroscopy can serve as a proof-of-principle experiment of the ONQ effect as well, as the side peaks can be detected if and only if the ONQ effect exists.

To analyze experimental feasibility of ONQ spectroscopy, we compare it quantitatively with Raman spectroscopy (Table III) due to the conceptual parallelism (Raman is phonon mode, i.e., nuclear positions, coupled to electronic orbitals, whereas ONQ is nuclear spin coupled to electronic orbitals, both leading to 2-photon process). We first examine the intensity of the Raman and the ONQ spectroscopy. The coupling strength of Raman spectroscopy per formula unit is determined by $f_{\text{Raman}} = g_{\text{Raman}} \mathcal{E}(\omega_{o1}) \mathcal{E}(-\omega_{o2})$, where $g_{\text{Raman}}$ is usually on the order of $10^5 \times \frac{2\pi \cdot \text{MHz}}{(\text{V/Å})^2}$. In contrast, the ONQ coupling strength is on the order of $10^2 \times \frac{2\pi \cdot \text{MHz}}{(\text{V/Å})^2}$ per formula unit, $10^3$ times smaller than $g_{\text{Raman}}$. This should not be a serious issue since the intensity of the side peaks is proportional to the number of atoms (nuclear spins) participating in the interaction. Notably, Raman spectroscopy has been demonstrated for characterizing the gas phase or atomically thin monolayer materials, which has a very limited number of atoms [44,45]. Hence, if the ONQ spectroscopy is used on a three-dimensional solid-state material, which have a large number of atoms within the optical interaction volume, then the intensity of the side peak would be significantly improved and should be detectable. Moreover, it is also possible to increase the intensity of the side peaks by using a stronger pumping laser or tip-enhanced or surface-enhanced Purcell factor, just like surface-enhanced Raman scattering (SERS). Besides, the low dissipation rate of the nuclear spins could also enhance the intensity of the side peaks [46].

Besides intensity, the linewidth of the side peaks is another important factor – the linewidth should be smaller than the frequency shift $\Delta_{ge}$, so that the side peaks can be spectrally resolvable. The linewidth of the side peaks originates in the linewidth of the pumping laser, the relaxation of nuclear spins, and the inhomogeneity broadening. Both the laser linewidth and the relaxation rate of nuclear spins can be kept below 1 kHz, far below $\Delta_{ge}$ (MHz to GHz). The inhomogeneity broadening could come from e.g., strain and temperature inhomogeneity. In most situations, the inhomogeneity broadening can be orders-of-magnitude smaller than $\Delta_{ge}$. For example, our ab initio calculation indicates that in many cases a 1% strain only leads to a less than 10% change in $\Delta_{ge}$. Therefore, the linewidth of the side peaks can be kept below the frequency shift $\Delta_{ge}$, and the side peaks can be experimentally detectable.

The spatial resolution of optical approaches can be much higher and is limited only by the diffraction limit (hundreds of nm for visible light). With tip-enhanced Purcell factor in a surface rastering setup, even super-resolution imaging might be possible like in super-resolution Raman imaging. However, the frequency shift between the side peaks and the main peak is on the order of MHz to GHz. Therefore, the spectral resolution required is relatively high compared with that in conventional spectroscopy techniques such as Raman spectroscopy, which can be a challenge in practice. Fortunately, Raman-like spectroscopy with spectral resolution down to the sub-kHz level has been demonstrated experimentally [46].

**Quantum memory.** Nuclear spins have the unique advantage of long coherence time even at room temperature [7,8] and are compelling candidates for quantum memory applications. In this section, we demonstrate that using the ONQ effect, the quantum information carried by optical photons can be directly stored in nuclear-spin quantum memory. To enhance the coupling rate between optical photons and nuclear spins, we consider a nuclear spin ensemble (NSE) with $N$ nuclear spins [47]. The quantum control over NSE with $N \sim 10^5$ has been realized using ancillary electron spins in, for example, quantum dot systems [48–50]. With the ONQ effect, a larger number of nuclear spins can be simultaneously controlled without the need for electron spins. The number density of nuclear spins in a pristine crystal is typically on the order of $\rho \sim 10^{28}$ m$^{-3}$, so a crystal with [1 μm]$^3$ size contains $N = 10^{10}$ nuclear spins. Note that the spot size and penetration depth of an off-resonance laser can be much greater than 1 μm, so the optical field in the [1 μm]$^3$ sample can be considered uniform. For NSE manipulation, the appropriate size of the crystal is discussed in the Appendix A.1.1. A pure crystal sample can have good homogeneity except on the surfaces, which can be advantageous for improving the coherence time of the quantum memory. The coherence time of the NSE can be further enhanced with quantum control techniques [47,51,52].

The ground state of the whole NSE is $|G\rangle \equiv |g_1 g_2 \cdots g_N\rangle$, in which all nuclear spins are on the ground state. The first collective excited mode of the NSE can be described as $|E\rangle \equiv \sum_{i=1}^{N} c_i |g_1 g_2 \cdots g_{i-1} e_i g_{i+1} \cdots g_N\rangle$, where the coefficients satisfy $\sum_{i=1}^{N} |c_i|^2 = 1$. One can see that on average one nuclear spin is on the excited state when the NSE is on $|E\rangle$. The energy splitting between $|G\rangle$ and $|E\rangle$ is denoted as $\Delta_{GE} = \epsilon_E - \epsilon_G$. In a uniform sample, one has $\Delta_{GE} \approx \Delta_{ge}$ when the interactions between nuclear spins are ignored. One challenge for the NSE qubit is the initialization to $|G\rangle$. In this regard, it can be advantageous to use nucleus with large quadrupolar energies, which can reach 1GHz, equivalent to 50 mK. Besides thermal cooling, the initialization of the NSE can also be facilitated by laser cooling using the ONQ effect [53]. Still, achieving a quantum memory in a NSE via the ONQ would be experimentally challenging due to the small energy scale of the nuclear spins compared to their thermal energy even at cryogenic temperature – the NSE needs to be cooled down to a very low temperature, and it is necessary to keep the temperature rise in the NSE due to laser illumination below several mK. The NSE is pumped with an $\omega_{o2}$-laser and is put into an optical cavity resonant with $\omega_{o1} = \omega_{o2} + \Delta_{GE}$. Assuming a uniform distribution (uniform sample under uniform laser), one has the collective ONQ Hamiltonian as $\mathcal{H} = G_o(|G\rangle\langle E| a_{o1}^+ + h.c.)$, where $a_{o1}^+$ is the creation operator of the $\omega_{o1}$-photon, and the collective ONQ coupling strength is $G_o = g_o \sqrt{N} \mathcal{E}(-\omega_{o2}) \mathcal{E}_{\text{zpf}}$, with $\mathcal{E}_{\text{zpf}} = \sqrt{\frac{\omega_{o1}}{\varepsilon_r \varepsilon_0 V_{o1}}}$ the zero point electric field of the $\omega_{o1}$-photon. $\varepsilon_0$ is the vacuum permittivity. $\varepsilon_r$ is the relative permittivity of the host material, and is taken as $\varepsilon_r = 5$ in the following. $V_{o1}$ is the mode volume of the optical cavity. If $\omega_{o1} = 1$ eV and $\rho \equiv \frac{N}{V_{o1}} = 10^{28}$ m$^{-3}$, then one has $G_o$ [MHz] $\sim 0.12 \times \mathcal{E}(\omega_{o2})$ [MV/cm], that is, one has $G_o = 60$ kHz when $\mathcal{E}(-\omega_{o2}) = 0.5$ MV/cm. Considering that the quality factor of an optical cavity can reach [54–56] $10^{10}$, the decay rate of the cavity photon is $\kappa_o \approx 24$ kHz, which is below the coupling strength $G_o$.

Next, we discuss the protocol of the NSE quantum memory. During the storage stage, the NSE is initialized to $|G\rangle$. When an $\omega_{o1}$-photon comes in, the NSE can be excited to $|E\rangle$ via the ONQ interaction, and thus memorize the incoming $\omega_{o1}$-photon. For readout, we propose to use the resonant emission of the $\omega_{o1}$-photon. Here, the NSE is again pumped with the $\omega_{o2}$-field, and the $\omega_{o1}$-cavity is tuned on-resonance with the ONQ transition. In this case, an $\omega_{o1}$-photon can be emitted if the NSE is on $|E\rangle$. In contrast, if the NSE is on $|G\rangle$, then the $\omega_{o1}$-photon cannot be emitted. One can thus detect whether the $\omega_{o1}$-photon is emitted to determine the state of the NSE. The emission rate of the $\omega_{o1}$-photon (if the NSE is on $|E\rangle$) is $\mathcal{R} = \frac{4G_o^2}{\kappa_o}$, which is 0.6 MHz when $\mathcal{E}(\omega_{o2}) = 0.5$ MV/cm and $\kappa_o = 24$ kHz. This is a relatively high emission rate, and should be detectable by current single-photon detectors [57,58]. It is also possible to readout using the dispersive interaction between the NSE and an off-resonant anharmonic cavity (Appendix A.2.2). Meanwhile, we would like to remark that $|E\rangle$ is the first excited state of the NSE, whereby one nuclear spin is on the excited state on average. It is possible to $N_n > 1$ nuclear spins on the excited states on average as well (similar to the multi-phonon state in Ref. [59]). In this case, the emission rate discussed above would be further enhanced by a factor of $N_n$. This could facilitate the experimental demonstration of the ONQ effect using NSE.

Finally, we would like to note that besides the desired storage/readout transitions as described above, it is also possible for the NSE to do other (undesired) transitions. For example, when the NSE is in $|G\rangle$ and is pumped with the $\omega_{o2}$-field, the NSE can spontaneously jump to $|E\rangle$ and emit a photon with frequency $\omega_{o2} - \Delta_{GE}$. However, the transition rate of such a process is strongly suppressed [60] by the $\omega_{o1}$-cavity by a factor of $r \sim \frac{\kappa_o^2}{4\Delta_{GE}^2 + \kappa_o^2}$. If $\Delta_{GE} = 2\pi \times 1$ GHz and $\kappa_o = 2\pi \times 1$ MHz, one has $r \sim 2.5 \times 10^{-7}$, which barely affects the fidelity of the nuclear-spin-based quantum memory.

**Quantum transduction.** We finally discuss another promising application of the ONQ effect, namely the quantum transduction between MW/RF and optical photons. We already described how the ONQ effect establishes an interface between optical photons and nuclear spins. On the other hand, the nuclear spins can also be coupled to MW/RF photons with a frequency $\omega_m$ through the Zeeman interaction. Therefore, nuclear spins can serve as the media for the transductions between MW/RF and optical photons. For brevity, we assume $\omega_{o1} > \omega_{o2}$ in the following. Treating the $\omega_{o1}$-laser as a classical pumping field, and second-quantizing the $\omega_{o2}$- and $\omega_m$-photons, we obtain the transduction Hamiltonian in the rotating-frame of $\omega_m$

$$H_{m \leftrightarrow o} = \sum_i \left\{ \frac{\delta}{2}(|e_i\rangle\langle e_i| - |g_i\rangle\langle g_i|) + g_o \mathcal{E}_{\text{zpf}} \mathcal{E}(\omega_{o1}) a_{o2}^+ |g_i\rangle\langle e_i| + g_m \mathcal{B}_{\text{zpf}} a_m^+ |g_i\rangle\langle e_i| + h.c. \right\} \quad (5)$$

Here $i$ labels each of the $N$ nuclear spins. The frequency matching condition requires $\omega_m = \omega_{o1} - \omega_{o2}$, while the nuclear spins are off-resonance with the MW/RF photon with a detuning $\delta \equiv |\omega_m - \Delta_{ge}|$. $\mathcal{E}_{\text{zpf}} = \sqrt{\frac{\omega_{o2}}{\varepsilon_r \varepsilon_0 V_{o2}}}$ is the zero-point electric field of the $\omega_{o2}$-photon as described before. $\mathcal{B}_{\text{zpf}} = \sqrt{\frac{\mu_0 \mu_r \omega_m}{V_m}}$ is the zero-point magnetic field of the $\omega_m$-photon. $\mu_o$ is the vacuum permeability, while the relative permeability of the host material is assumed to be $\mu_r = 1$. In the large detuning limit, the effective transduction Hamiltonian becomes

$$H_{\text{eff}} = G_o a_{o2}^+ |g\rangle\langle e| + G_m a_m^+ |g\rangle\langle e| + h.c. \quad (6)$$

where $G_o = g_o \sqrt{N} \mathcal{E}(\omega_{o1}) \mathcal{E}_{\text{zpf}}$ and $G_m = g_m \sqrt{N} \mathcal{B}_{\text{zpf}}$ are the total coupling rate of the NSE with optical and MW/RF fields, respectively. To estimate the magnitude of $G_o$ and $G_m$, we assume $\omega_{o2} = 1$ eV and $\omega_m = 2\pi \cdot 1$ GHz. We also assume $N = \rho V_{o2}$, where $\rho \sim 10^{28}$ m$^{-3}$ is the number density of the nuclear spins. Thus, one has $G_o \approx 0.24$ MHz when $\mathcal{E}(\omega_{o1}) = 2$ MV/cm. We also assume the quality factor of the $\omega_{o2}$-cavity is $Q = 10^{10}$. Note that one can use a weaker pumping field $\mathcal{E}(\omega_{o1})$ if $Q$ is higher ($Q > 10^{10}$ has been achieved [54–56]). For $G_m$, we further assume an MW/RF mode volume of $V_m = 1$ mm$^3$, although a smaller MW mode volume is achievable [61]. Then one has $G_m \sim 0.3$ MHz when $N \sim 10^{18}$, corresponding to a crystal size of 0.1 mm$^3$. We assume the quality factor of the MW/RF cavity is $Q = 10^5$ [62].

In the large detuning regime ($\delta \gg G_o, G_m$), the nuclear spin modes can be further adiabatically eliminated [63–65] and one can reach a linear coupling between the optical and MW/RF modes $H_{\text{eff}} = G_{om}(a_{o2}^+ a_m + h.c.)$, with $G_{om} \equiv \frac{G_o G_m}{\delta}$. However, this approach ignores the relaxation of the transducer (nuclear spins in the current situation). To clearly show how dissipations in the system can affect the transduction process, we retain the complete form Eq. (6) and introduce dissipation due to the optical and MW/RF cavity (with rates $\kappa_{o2}$ and $\kappa_m$, respectively) as well as the nuclear spin relaxation (with rate $\Gamma_n$). The fidelity of the quantum transduction can be estimated by solving the master equation

$$\frac{d\rho}{dt} = -i[H_{\text{eff}}, \rho] + \frac{1}{2}\kappa_{o2}\zeta(a_{o2}) + \frac{1}{2}\Gamma_n\zeta(\sigma^-) + \frac{1}{2}\kappa_m\zeta(a_m), \tag{7}$$

where $\rho$ is the density matrix of the total system, $\zeta(o) = 2o\rho o^+ - o^+o\rho - \rho o^+o$ is the Lindblad operator for a given operator $o$, such as $\sigma^- \equiv |g\rangle\langle e|$. The master equation simulations are performed using the QuTiP package [66,67]. In the simulations, $\kappa_{o2}$ and $\kappa_m$ are determined by the quality factor of the cavities, which is taken as $10^{10}$ ($10^5$) for the optical (MW/RF) cavity, respectively, while even higher quality factors have been achieved in experiments [54–56,62]. Note that when cavities with higher quality factors are used, the coupling strength $G_o$ and $G_m$ required to achieve efficient transduction will be smaller.

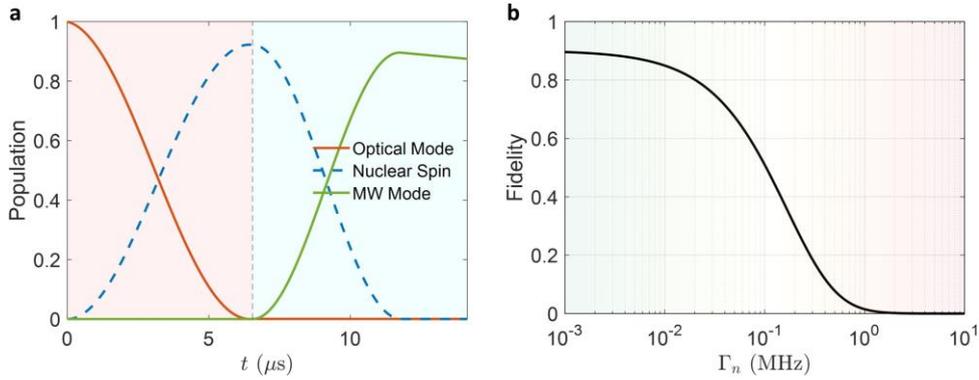

**Figure 3 Simulation results of optical to MW/RF transduction. (a)** Population of different subsystems as a function of time $t$. Here we take $\Gamma_n = 1$ kHz. In the light-red shaded region, the optical-nuclear interaction is turned on, while in the light-blue shaded region, the MW-nuclear interaction is turned on. **(b)** Fidelity of the optical to MW transition as a function of NSE relaxation rate $\Gamma_n$. In the simulations we take $G_o \approx 0.24$ MHz and $G_m \sim 0.3$ MHz.

We consider a simple transduction protocol using sequential swap gates [68,69]. We will use optical to MW/RF transduction as an example, whereas the reverse MW/RF to optical to transduction can be analyzed similarly. Specifically, an incoming optical photon is stored in the cavity and interacts with the nuclear spins ($a_m^+|g\rangle\langle e|$) for a certain amount of time corresponding to an effective $\pi$ pulse, while the MW/RF cavity is detuned. After the population is transferred to nuclear spins, the MW/RF interaction ($a_{o2}^+|g\rangle\langle e|$) is turned on, while the optical pumping is turned off. Then the population can be transferred from nuclear spins to MW/RF photons. In Figure 3a we show that this simple swap protocol can yield a transduction fidelity ~90% with parameter settings described above. Notably, the relaxation of transducers (nuclear spins) strongly affects the transduction fidelity. From Figure 3b, one can see that the transduction fidelity significantly deteriorates when $\Gamma_n \gtrsim 0.1$ MHz. In comparison with other methods for MW/RF-optical transduction [70] using e.g., phonons [71–74] or electron spins [63,65,82,69,75–81] as transducers, a unique advantage of using the ONQ effect is that the nuclear spins have a long lifetime and good immunity to noise in the environment, thus high fidelity can be achievable at relatively high temperatures. Indeed, $\Gamma_n$ can be as low as Hz even at elevated temperatures [83,84]. The transduction based on the ONQ effect also benefits from the enormous number density of nuclear spins ($\sim 10^{28}$ m$^{-3}$) in pure solid-state systems. Therefore even though the coupling strength between a single nuclear spin and MW/RF/optical photons is much weaker than that of electron spins, the sample size required here is only millimeter size, similar to that in Ref. [63].

**Discussions.** Before conclusion, we would like to discuss some issues relevant to the experimental demonstration of the ONQ effect. Some specific issues, such as the proper size of the crystal sample for NSE, and the temperature rise due to laser illumination, are discussed in Appendix A. Here, we first discuss the proof-of-principle demonstration of the ONQ effect. As discussed before, the manipulation of the single nuclear spin or the nuclear spin ensemble can serve as the proof-of-principle experiments of the ONQ effect. These two experiments are closely connected to the quantum nature of the nuclear spin (ensemble). In contrast, the ONQ spectroscopy (detecting the side peaks due to the ONQ scatterings) can be considered as a more "classical" proof-of-principle experiment and could be easier to demonstrate. Besides, one can also use techniques developed in NMR technology to detect the nuclear spin dynamics induced by the ONQ effect. This is also a relatively easier experiment, which has been used to demonstrate the NER [85].

To demonstrate the ONQ effect, the linewidth of the laser/photon should be kept small, preferably on the order of MHz to kHz. This is also relevant to the necessity of optical cavities with high quality factors, or spectrometers with high spectral resolution. This is because the energy scales of nuclear spins, including resonance frequency, Rabi frequency, linewidth, etc. are usually very small (kHz to MHz, at most GHz). Hence, the linewidth of the laser/photon should be kept small as well, so that the nuclear spin dynamics would not be deteriorated (see Appendix A.3.2 for detailed discussions). Fortunately, laser sources with small linewidth down to the sub-Hz level [86–90] and optical cavities with high $Q$ factor up to $10^{10}$ and above [54–56] have been realized in experiments. Another issue is the appropriate temperature for experimentally demonstrating the ONQ effect. Pertinent to the ONQ effect itself, the temperature is relevant in that nuclear spins could have longer relaxation/decoherence time at lower temperature. Meanwhile, low temperature could improve the experimental results. For example, the signal-noise ratio of certain detectors could be higher at lower temperatures. Nevertheless, low temperature is not a necessary condition for the ONQ effect itself. Furthermore, the laser illumination could result in heating effect and sample damage. This limits the maximum laser intensity and also the maximum achievable ONQ Rabi frequency (tens of kHz for single nuclear spin and MHz for NSE, see Appendix A.1).

As we discussed before, the ONQ effect does not require electron spins $S \neq 0$, which is a unique advantage. If $S \neq 0$, then there will be hyperfine interactions $H_{\mathrm{hf}} = \sum_{ij} A_{ij} I_i S_j$ between electron and nuclear spins. Similar to the quadrupole tensor $Q$, the hyperfine tensor $A$ is also determined by the electron wavefunctions. Hence, under two-color photons $\omega_{o1}$ and $\omega_{o2}$, $A$ would oscillate with the frequency $|\omega_{o1} - \omega_{o2}|$, which can be in resonance with either electron or nuclear spin frequencies. This provides another degree of freedom to manipulate the hybrid electron-nuclear spins system.

In summary, we propose the ONQ effect which can serve as an efficient interface between optical photons and nuclear spins. As electron spins are not required and the frequencies of the optical photons can be arbitrary, the ONQ effect provides substantial flexibilities that could empower various promising applications ranging from isotope spectroscopy to quantum technologies, including quantum control, quantum memory, and quantum transduction between optical and MW/RF fields as required in quantum communication.

## Appendix A: Experimental Considerations

In this section, we discuss several issues relevant to the experimental realization of the opto-nuclear quadrupolar (ONQ) effect.

### 1. Heating Under Laser Illumination

The ONQ effect requires laser fields with electric field strength on the order of $0.1 \sim 1\ \mathrm{MV/cm}$. In this section, we show that the material sample should be able to sustain such a laser illumination. This is particularly because the laser frequency is below the bandgap of the material, so that the direct one-photon absorption of the laser energy is minimal. In the following, we first suggest the size of the crystal sample of the nuclear spin ensemble (NSE) for demonstrating the ONQ effect. Then, we show that the temperature rise in the electron/phonon system due to the laser illumination is only on the order of 1~10 K when the electric field is 1 MV/cm, so the sample damage due to heating effect should be minimal.

## 1.1 Proper Size of the Crystal Sample

As discussed in the main text, for a single nuclear spin, the ONQ coupling strength is $f = g_o \mathcal{E}(\omega_{o1})\mathcal{E}(-\omega_{o2})$. Hence, it is desirable to have strong electric fields $\mathcal{E}(\omega_{o1})$ and $\mathcal{E}(-\omega_{o2})$ (below possible sample damage threshold). Meanwhile, for a NSE in an optical cavity, the ONQ coupling strength is $f = g_o \mathcal{E}(-\omega_{o2})\sqrt{\frac{\omega_{o1} N}{\epsilon_r \epsilon_0 V_{o1}}}$ with $N$ the number of nuclear spins and $V_{o1}$ the mode volume of the $\omega_{o1}$-cavity. Hence, it is desirable to have strong $\mathcal{E}(-\omega_{o2})$ as well. On the other hand, $\rho = \frac{N}{V_{o1}}$ is approximately the number density of nuclear spins, and is independent of the size of the crystal sample, which should be comparable with $V_{o1}$. The optical fields can have stronger intensity when focused onto a smaller spot-size. The spot-size is limited by the wavelength of the optical field, which is on the order of 1 μm. To this end, we suggest using a crystal sample with transverse area of $[1\mu m]^2$ to demonstrate the ONQ effect in an NSE. Note that this is not a necessary condition, and smaller samples can be used as well, especially for single nuclear spin manipulation.

Another factor that should be considered is the penetration depth $d_p$ of the optical field. The depth $d$ of the sample should be much smaller than $d_p$ to ensure that the light field in the sample is uniform. Since we propose to use optical fields whose frequencies $\omega_{o1\,(o2)}$ are below the bandgap $E_g$ of the crystal, the absorption of the laser energy via the direct one-photon process is in principle zero. Meanwhile, when $2\omega_{o1\,(o2)} > E_g$, the absorption of the laser can result from the the two-photon process, whereby electrons do interband transitions by simultaneously absorbing two photons. The intensity of the two-photon process is determined by the two-photon absorption coefficient $\beta$, which is on the order of $10^{-10}$ m/W in typical semiconductors [91–95]. The penetration depth of an optical field due to the two-photon absorption is

$$d_{\mathrm{p}} = \frac{1}{\beta P_{\mathrm{in}}}$$
$$= \frac{2}{\beta \varepsilon_0 c_0 \mathcal{E}^2}, \quad \text{(A1)}$$

where $P_{\mathrm{in}} = \frac{1}{2} c_0 \varepsilon_0 \mathcal{E}^2$ is the incident laser power. Here $c_0$ is the speed of light. If one uses $\mathcal{E} = 1$ MV/cm, $\beta = 10^{-10}$ m/W, then one has $d_{\mathrm{p}} \approx 750$ μm. Hence, if the depth of the crystal is $d \approx 0.1$ μm, then the optical field inside can be considered as uniform, since $d \ll d_{\mathrm{p}}$.

In summary, for NSE manipulation, we suggest using a crystal sample with a $1 \times 1 \times 0.1$ μm$^3$ dimension, which can ensure that the optical fields inside are uniform. Considering that the number density of nuclear spins can be order of $10^{28}$ m$^{-3}$ in typical semiconductors (e.g., zinc blende GaAs), the $1 \times 1 \times 0.1$ μm$^3$ crystal sample contains around $10^9$ nuclear spins.

### 1.2 Temperature Rise in the Electron/Phonon System

Next, we estimate the temperature rise of the crystal sample under laser illumination. The duration of the laser illumination $\tau_{\mathrm{laser}}$ should be on the order of μs to ms, since the ONQ Rabi frequency is on the order of tens of kHz to MHz. On the other hand, the timescale of electron/phonon dynamics $\tau_{e,p}$ is on the order of fs to ps. Since $\tau_{\mathrm{laser}} \gg \tau_{e,p}$, the laser can be treated as a continuous wave laser to estimate the temperature rise in the electron/phonon system.

The laser energy absorption power per unit area can be estimated with [96]

$$P_{\mathrm{abs}} = P_{\mathrm{in}} \left( 1 - e^{-\frac{d}{d_{\mathrm{p}}}} \right)$$
$$\approx P_{\mathrm{in}} \frac{d}{d_{\mathrm{p}}}. \quad \text{(A2)}$$

Then, the temperature rise in the electron/phonon system can be estimated from

$$\Delta T_{ep} = \frac{P_{\mathrm{abs}}}{k_{\mathrm{th}}} d$$
$$= \frac{1}{2} \frac{c \epsilon_0 \mathcal{E}^2}{d_{\mathrm{p}}} \frac{d^2}{k_{\mathrm{th}}}. \quad \text{(A3)}$$

if we use $\mathcal{E} = 1$ MV/cm ($d_p = 750$ μm), $d = 0.1$ μm, and take the thermal conductivity as $k_{\mathrm{th}} = 10$ W·m$^{-1}$·K$^{-1}$, then one has $\Delta T_{ep} = 15$ K, which is mild and usually would not lead to any damage in the crystal sample. If we use $\mathcal{E} = 0.1$ MV/cm ($d_p = 7.5 \times 10^4$ μm), then one has $\Delta T_{ep} = 1.5$ mK.

We would like to remark that it is crucial to forbid the one-photon process. If one photon absorption is allowed, then one has $d_p \approx \frac{\lambda}{2\pi\varepsilon^{(1)}}$, where $\varepsilon^{(1)}$ is the imaginary part of the dielectric constant and can be on the order of 10 ~ 100 for an above-bandgap laser. The temperature rise due to one-photon absorption could be $10^5 \sim 10^6$ times larger than that due to two-photon absorption, and reach thousands of Kelvins with $\mathcal{E} = 1$ MV/cm. This would damage the sample. Consequently, high-quality crystal samples are desirable, because unwanted defects in the crystal sample could result in-gap states and lead to one-photon absorption of laser energy. On the other hand, when certain defects (such as NV center) are required, the laser frequency should be kept below the electronic transition energy of the defect states (e.g., 637 nm for NV$^{-1}$).

Moreover, if the laser frequency satisfies $2\omega_{o1\,(o2)} < E_g$, then the two-photon absorption is forbidden as well, and the leading order contribution to the absorption would be the three-photon process. In this case, the laser absorption and the temperature rise could be even smaller than those estimated in Eqs. (A2, A3).

**1.3 Tunnelling Ionization**

In the previous section, we discussed the absorption of laser energy due to the two-photon process, which corresponds to the multi-photon ionization (MPI). Besides, ionization can also result from the tunnelling effect. Notably, the relative intensity of the multi-photon and tunnelling ionization is determined by the Keldysh parameter [97,98]

$$\gamma = \frac{\omega}{e}\left[\frac{mc_0 n\epsilon_0 E_g}{P_{\text{in}}}\right]^2, \tag{A4}$$

where $m$ is the electron mass, $n$ is the refraction index. When $\gamma \gtrsim 1.5$, the tunnelling process is much weaker than the multi-photon process. Putting $\mathcal{E} = 1$ MV/cm, $\omega = 2$ eV and $E_g = 2.2$ eV in the equation above, we obtain $\gamma \sim 100$. This indicates that the multi-photon process dominates the absorption of the laser energy, and the tunnelling ionization is negligible. This further corroborates that the material sample should be able to sustain the 1 MV/cm field.

## 2. Readout of Nuclear Spin States

In the main text, we discussed possible approaches for the readout of the quantum state of the nuclear spins. In this section, we propose alternative approaches for the readout of both a single nuclear spin and an NSE.

### 2.1 Single Nuclear spin

In the main text, we mentioned that the single nuclear spin can be read out using ancillary electron spins. In the long term, it might be desirable to totally get rid of electron spins and to develop an all-optical control over nuclear spins. One possible approach is to use high-quality optical cavities and high-efficiency single photon detectors. A single nuclear spin is put into an optical cavity resonant with the $\omega_{o1}$-photon and is pumped with a $\omega_{o2}$-laser ($\omega_{o1} > \omega_{o2}$). After the second-quantization of the $\omega_{o1}$-photon, the ONQ coupling would be

$$H_{\text{ONQ}} = g_o \mathcal{E}(-\omega_{o2}) \mathcal{E}_{\text{zpf}}^{o1} (|g\rangle\langle e| a_{o1}^+ + h.c.), \tag{A5}$$

where $h.c.$ stands for Hermitian conjugate, $a_{o1}^+$ is the creation operator of the $\omega_{o1}$-photon, and $\mathcal{E}_{\text{zpf}}^{o1} = \sqrt{\frac{\omega_{o1}}{\epsilon_r \epsilon_0 V_{o1}}}$ is the zero-point field of the $\omega_{o1}$-photon in the cavity with $V_{o1}$ the mode volume. When the nuclear spin is on excited state $|e\rangle$, then it can emit a $\omega_{o1}$-photon via the ONQ effect and jump back to $|g\rangle$. On the other hand, if the nuclear spin is on ground state $|g\rangle$, then the $\omega_{o1}$-photon would not be emitted. Hence, the state of the nuclear spin can be determined by detecting whether the $\omega_{o1}$-photon is emitted using a single photon detector. The emission rate of the $\omega_{o1}$-photon is [60]

$$\begin{aligned} R &= \frac{2\left[g_o \mathcal{E}(-\omega_{o2}) \mathcal{E}_{\text{zpf}}^{o1}\right]^2}{\kappa_{o1}} \\ &= \frac{2[g_o \mathcal{E}(-\omega_{o2})]^2}{\epsilon_r \epsilon_0 V_{o1}} Q_{o1}, \end{aligned} \tag{A6}$$

where $\kappa_{o1}$ is the cavity decay rate, and $Q_{o1} \equiv \frac{\omega_{o1}}{\kappa_{o1}}$ is the quality factor of the cavity. One can see that the emission rate $R$ can be faster for a cavity with large quality factor and small mode volume.

In practice, optical cavities with quality factor above $10^{10}$ has been demonstrated [56,99], and the mode volume can be down to $10^{-22}$ m$^3$ by nano-photonics design [100]. Using $Q_{o1} = 10^{10}$ and $V_{o1} = 10^{-22}$ m$^{-3}$, one has $R$ [Hz] $\approx 12 \times [\mathcal{E}(-\omega_{o2})]^2$ [MV/cm]$^2$. That is, one has $R = 60$ Hz when $\mathcal{E}(-\omega_{o2}) = 5$ MV/cm. This is a relatively small emission rate, and thus high efficiency single photon detectors would be desired [57,58]. Besides, it also requires an optical cavity that simultaneous has high $Q$-factor and small mode volume. This could be challenging as well. Hence, we consider this all-optical readout of single nuclear spin as a long-term goal, which could be facilitated by the development of quantum/classical photonics.

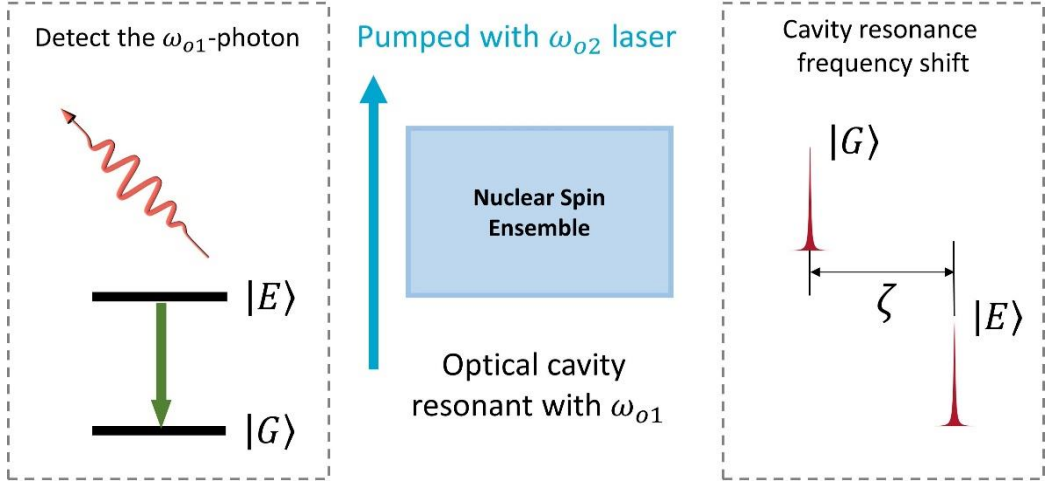

**Figure 4:** Readout the quantum state of the NSE. Left box: the cavity is on-resonance with the ONQ transition, and one can detect whether the $\omega_{o1}$-photon is emitted to determine the state of the NSE. Right box: an anharmonic cavity is off-resonance with the ONQ transition. One can detect the shift in the resonance frequency of the cavity to determine the quantum state of the NSE.

## 2.2 Nuclear Spin Ensemble

In the main text, we showed that the quantum state of the NSE can be readout by detecting the emission of optical photons in a resonant cavity (left box in Figure 4). Here we propose another approach for the non-demolition measurement of the quantum state of the NSE, using the dispersive interaction [59,101–103] with an off-resonance anharmonic optical cavity. In this case, the $\omega_{o1}$-cavity is tuned off-resonance with the ONQ transition, i.e., $\delta \equiv \omega_{o1} - \omega_{o2} - \Delta_{GE} \neq 0$. Meanwhile, the cavity has an anharmonicity $\alpha$, which could result from e.g., the interaction with an ancillary atom [104,105] and can reach above 1 MHz. In this case, the resonance frequency of the cavity depends on the states of the NSE (right box in Figure 4), and the shift in the resonance frequency is given by

$$\zeta = \frac{2\left[g_o\sqrt{N}\mathcal{E}(-\omega_{o2})\mathcal{E}_{\text{zpf}}^{o1}\right]^2}{\delta} \frac{1}{1+\delta/\alpha}. \quad (A7)$$

One has $\zeta \approx 30$ kHz when $\mathcal{E}(-\omega_{o2}) = 0.5$ MV/cm, $\delta = 0.2$ MHz, and $\alpha = 1$ MHz. Such a $\zeta$ is resolvable considering that the linewidth of the cavity is around 24 kHz when $Q = 10^{10}$. Therefore, by detecting the shift in the resonance frequency of the cavity, one can indirectly probe the NSE state. This approach for detecting NSE state could be more challenging than the *resonant photon emission* approach described in the main text, since introducing anharmonicity in the optical cavity could affect the properties of the NSE (e.g., coherence time) as well. Hence, the system needs to be carefully designed, which we leave as a future work.

## 3. Effect of Finite Linewidth

In this section, we discuss the influence of the finite linewidth of electrons and lasers/photons. We will show that the linewidth of the electron makes negligible influence. Meanwhile, the linewidth of laser/photon plays a relatively important role, and should be kept below the Rabi frequency of the ONQ nuclear spins transitions.

### 3.1 Electronic Transition Linewidth

The electron dynamics is usually very fast compared with NMR frequency, and the electron linewidth is usually $\eta \lesssim 1$ meV, equivalent to an electron lifetime from picosecond to sub-nanosecond [39–41]. However, as can be observed from Eq. (4) in the main text, the electron linewidth $\eta$ makes an important role only when $\omega_{o1}$ (or $\omega_{o2}$) is very close to the electronic transition energy $E_{mn}$. As discussed before, we propose to use $\omega_{o1\,(o2)} < E_g$ with $E_g$ the bandgap, so $\omega_{o1}$ (or $\omega_{o2}$) would not be close to $E_{mn}$, since one has $\omega_{o1\,(o2)} < E_g \leq E_{mn}$. Meanwhile, the NMR frequency $\omega_m \equiv \omega_{o1} - \omega_{o2}$ does not play an important role as well, because $\omega_m$ is on the order of MHz to GHz and is too small compared with $E_{mn}$ or $\omega_{o1\,(o2)}$, which are on the order of $10^{15}$ Hz. Moreover, the NER effect, which uses microwave/radio frequency electric field to modulate the EFG generated by electrons, has already been demonstrated experimentally [16,85]. This corroborates that the electron linewidth does not need to be smaller than the NMR frequency.

### 3.2 Laser/Photon Linewidth

The laser/photon linewidth is an important parameter. Before further discussions, we would like to remark that while it is desirable to have a small laser/photon linewidth, it is *not necessary* to have a laser/photon linewidth smaller than the NMR frequency. Nonlinear optical responses can exist even if the laser/photon linewidth is wide. For example, the bulk photovoltaic effect [106], whereby a DC charge current is generated under light illumination, can happen under white light [107], which has an ultra-large "linewidth".

As can be observed in Eq. (A6), the ONQ transition rate is inversely proportional to the photon linewidth. Here we further examine the importance of the laser/photon linewidth. We assume the laser/photon has a Lorentzian line shape, and the two-color laser/photon used in the ONQ effect are described by

$$\mathcal{E}_1(\omega_1) = \frac{1}{\pi}\frac{\kappa_{o1}}{(\omega_1 - \omega_{o1})^2 + \kappa_{o1}^2},$$
$$\mathcal{E}_2(\omega_2) = \frac{1}{\pi}\frac{\kappa_{o2}}{(\omega_1 - \omega_{o2})^2 + \kappa_{o2}^2}. \quad (A8)$$

That is, the central frequency is $\omega_{o1}$ ($\omega_{o2}$) and the linewidth is $\kappa_{o1}$ ($\kappa_{o2}$).

When two-color fields with frequencies $\omega_1$ and $\omega_2$ are combined to trigger the Rabi oscillation of the nuclear spins, the efficiency of the Rabi oscillation is proportional to $\eta(\omega_1, \omega_2) = \frac{f_{\text{Rabi}}^2}{f_{\text{Rabi}}^2 + (\omega_1 - \omega_2 - \Delta)^2}$, where $\Delta$ is the nuclear spin resonance frequency and $f_{\text{Rabi}}$ is Rabi frequency. Under laser/photon with line shapes described by Eq. (A8), the overall efficiency of the Rabi oscillation can be estimated with

$$\eta_0 = \iint d\omega_1 d\omega_2 \mathcal{E}_1(\omega_1)\mathcal{E}_2(\omega_2)\eta(\omega_1, \omega_2)$$
$$= \frac{f_{\text{Rabi}}(f_{\text{Rabi}} + \kappa_{o1} + \kappa_{o2})}{(\omega_1 - \omega_2 - \Delta)^2 + (f_{\text{Rabi}} + \kappa_{o1} + \kappa_{o2})^2} \quad (A9)$$

One can see that when $\omega_1 - \omega_2 = \Delta$ and $\kappa_{o1} = \kappa_{o2} = 0$, the efficiency is $\eta_0 = 1$, as expected. Besides, one has $\eta_0 \sim 1$ when

$$|\omega_1 - \omega_2 - \Delta| \lesssim f_{\text{Rabi}}$$
$$\kappa_{o1}, \kappa_{o2} \lesssim f_{\text{Rabi}} \quad (A10)$$

That is, the detuning from perfect resonance frequency and the linewidth of the two-color laser/photon should be kept below the Rabi frequency $f_{\text{Rabi}}$, which can be on the order of kHz to MHz.

In practice, using some laser stabilizing techniques [86–88,108], the linewidth of the lasers can be relatively easily kept below kHz, and down to the sub-Hz level [89,90].

## 4. Effect of Finite Wavevectors

In the main text, we treat the laser fields as a spatially uniform field. In practice, the spatial pattern of the laser field is characterized by a finite wavevector $k$. That is, the optical field has a phase factor $e^{ik \cdot r}$, where $r$ is the position. Since $k$ is usually much smaller than the size of the Brillouin zone in typical materials, usually one uses $k \approx 0$ when theoretically studying optical processes (see e.g., Ref. [109]). Recently, we have also studied the finite-$k$ effect in nonlinear optical response [110], and find that it is very small for an optical field with $k \sim 1\ \mu m^{-1}$. In practice, the finite $k$ effect could result in the phase matching issue in nonlinear optical processes. But since our sample has only μm dimension, the phase matching efficiency should be very high [111].

## 5. Enhancing the Nuclear Quadrupolar Interaction

For certain applications, it could be desirable to have a strong nuclear quadrupolar interaction, which leads to large nuclear spin splitting. One approach for enhancing the nuclear quadrupolar interaction is to use isotopes with large quadrupolar moment. On the other hand, the nuclear quadrupolar interaction can be enhanced when the EFG is large. Usually, the EFG can be stronger if the symmetry of the lattice is lower. For example, in zinc-blende GaAs (space group $F\bar{4}3m$, no. 216), the EFG is zero because of the tetrahedral symmetry. In contrast, in wurtizte GaAs (space group $P6_3mc$, no. 186), the EFG is not zero because the symmetry is lower. Hence, to increase the EFG, a generic approach is to find materials with low symmetry, such as $HfO_2$ (space group $P2_1/c$, no. 14).

Meanwhile, the EFG tensor can be further enhanced by

- **Strain**. Strain can usually enhance the asymmetry and hence the magnitude of the EFG tensor $\mathcal{V}$. As an example, we apply uniaxial strain along the $z$ direction (crystallographic $c$-axis) of wurtzite GaN. One can see that a strain of 1% can improve $\mathcal{V}_{zz}$ by more than 100% in this case (Figure 5).
- **Point defect**. Point defect can usually significantly alter the local symmetry. An efficient approach is to introduce vacancies. For example, in diamond, if a nitrogen (N) simply substitutes a carbon atom, then the EFG tensor at the site of the N nuclei is still zero because of the tetrahedral symmetry. In contrast, if the N atom is associated with a vacancy (NV center), then the local symmetry around the N nuclei would be lower, and the EFG tensor would be non-zero, inducing a quadrupole splitting of the N nuclear spin.

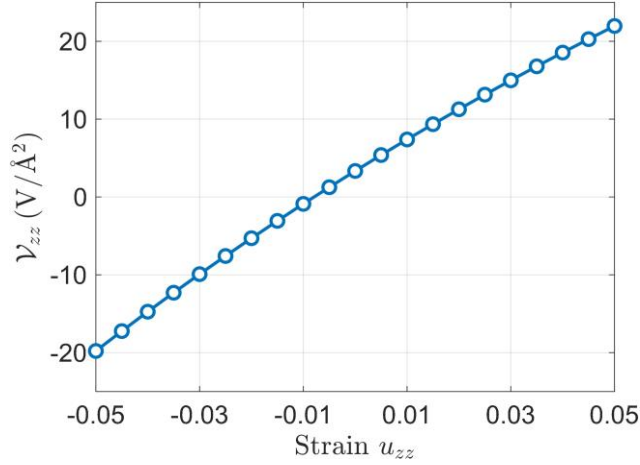

**Figure 5:** The $\mathcal{V}_{zz}$ component of the EFG at the site of Ga nuclei in wurtzite GaN as a function of uniaxial strain along $z$-direction.

Both strain and defects can be introduced either during or after the growth of the crystal, so that the EFG can be improved.

## Appendix B: Theoretical Approaches

In this section, we provide detailed discussions on the theoretical approaches for estimating the ONQ $\mathcal{D}$ tensor. We will show that the $\mathcal{D}$ we obtained should have the correct order of magnitude.

### 1. DFT Calculations of the $\mathcal{D}$ Tensor

The DFT calculations are performed with the Quantum Espresso [112,113] package. The exchange-correlation interactions are treated with the generalized gradient approximation (GGA) in the form of Perdew-Burke-Ernzerhof (PBE) [114]. Core and valence electrons are treated by the projector augmented wave (PAW) method [115] and plane-wave basis functions, respectively. The first Brillouin zone is sampled by a $\boldsymbol{k}$-mesh with a grid density of at least $2\pi \times 0.02$ Å$^{-1}$ along each dimension. For the calculation of the $\mathcal{D}$ tensor, we apply an electric field $\mathcal{E}$ in the simulation using the modern theory of polarization [116,117], and obtain the field gradient (EFG) tensor $\mathcal{V}$ as a function of $\mathcal{E}$. The EFG tensor is calculated using GIPAW [118], which is a component of the Quantum Espresso package. Then, the response $\mathcal{D}$ tensor is

$$\begin{aligned}\mathcal{D}_{ij}^{pq} &= \left.\frac{\partial^2 Q_{ij}}{\partial \mathcal{E}_p \partial \mathcal{E}_q}\right|_{\mathcal{E}=0} \\ &= \frac{eq}{2I(2I-1)}\left.\frac{\partial^2 \mathcal{V}_{ij}}{\partial \mathcal{E}_p \partial \mathcal{E}_q}\right|_{\mathcal{E}=0}\end{aligned} \quad (B1)$$

where $\frac{\partial^2 \mathcal{V}_{ij}}{\partial \mathcal{E}_p \partial \mathcal{E}_q}$ is obtained by fitting the $\mathcal{V}$-$\mathcal{E}$ curve obtained from DFT calculations.

## 2. Validity of the Theoretical Predictions on the $\mathcal{D}$ Tensor

In the following, we demonstrate the validity of our theoretical prediction of the magnitude of the $\mathcal{D}$ tensor, which characterizes the ONQ effect. In particular, we will show that $\mathcal{D}$ we obtained should have the correct order of magnitude.

### 2.1 DFT Calculations of Spin-Related Quantities

First, we would like to point out that the validity of DFT calculations of spin-related quantities for qubit research, including the quadrupole interaction, the hyperfine interaction, and the zero-field splitting, etc., has been systematically analyzed in Refs. [119,120]. It is found that DFT calculations usually agree well with experimental results in terms of ground state properties (i.e., no electric field). As an example, Ref. [121] studied the quadrupole splitting $C_q \equiv \frac{eq\mathcal{V}_{zz}}{h}$ of nitrogen-vacancy (NV) center using both DFT calculations and experimental measurements. Here $\mathcal{V}_{zz}$ is the largest principal value of the EFG tensor $\mathcal{V}$. It is found that DFT and experimental results are off by around 5 % (Table 1 therein).

Our own DFT calculations for the quadrupole splitting of NV⁻ is $C_q \approx 5.1$ MHz, in good agreement with experimental results of 4.95 MHz. Moreover, recently we have studied the temperature and strain dependence of spin-related quantities in NV centers [122,123]. The theoretical predictions agree well with experimental results as well.

### 2.2 The First-Order Response

Next, we study the first-order response of the quadrupole interaction to electric field, which corresponds to the nuclear electric resonance (NER) and the linear quadrupole Stark effect (LQSE). We will show that the theoretical predictions from both perturbation theory and DFT calculations give the correct order of magnitude of the strength of NER/LQSE, as compared with experimental results (Table IV)

Basically, the NER/LQSE can be described by the response function

$$\mathcal{C}_{ij}^{p} \equiv \frac{\partial Q_{ij}}{\partial \mathcal{E}_p} \tag{B2}$$

which describes how strongly the quadrupole interaction $Q$ changes with electric field $\mathcal{E}$. For the NER, a radio frequency electric field drives the Rabi oscillation between different nuclear spin states [16,18,85], and the Rabi frequency is approximately $f_{\text{Rabi}} \approx \mathcal{C}_{\text{off}}\mathcal{E}$, where $\mathcal{C}_{\text{off}}$ is the off-diagonal term of the $\mathcal{C}$ tensor in the basis of nuclear spin eigenstates. For the LQSE, an electric field leads to the shift in the nuclear spin transition frequency, $\delta\omega \approx \mathcal{C}_{\text{diag}}\mathcal{E}$, where $\mathcal{C}_{\text{diag}}$ is the diagonal term of the $\mathcal{C}$ tensors [124,125]. Using these relationships, the magnitude of the $\mathcal{C}$ tensor can be obtained from experimental data.

**Table IV:** Theoretical predictions and experimental results of the strength of the NER response (in $\frac{2\pi \cdot \text{MHz}}{\text{V/Å}}$)

| System | Response | Experiment | Perturbation Theory | DFT |
|---|---|---|---|---|
| $^{75}$As in zinc blende GaAs | NER | 20 [85] | 9 | 5 |
| $^{123}$Sb defect in Si | NER | 30 [16] | 8 | 12 |
| $^{69}$Ga in zinc blende GaAs | LQSE | 35 [124] | 7 | 3 |
| $^{35}$Cl in CCl$_4$ | LQSE | 4.8 [125] | 0.7 | 0.5 |

**Experimental Results.** There are several experimental works on the NER or the LQSE responses. We estimate the strength of the responses from data therein, which are listed in the third column of Table IV.

**Perturbation Theory Estimation.** The $\mathcal{C}$ tensor can be obtained from first-order perturbation theory. In the single-particle approximation, one has [18,126]

$$\mathcal{C}_{ij}^{p}(\omega) = \frac{e^2 q}{2I(2I-1)} \sum_{mn} \frac{f_{nm}[\mathcal{V}_{ij}]_{nm}[r_p]_{nm}}{E_{mn} - \omega + i\eta} \tag{B3}$$

The meaning of each term is defined in the main text around Eq. (4). Notably, here $\omega$ should be comparable with nuclear spin energies.

In typical semiconductors, one has $E_{mn} \sim 1$ eV, $\eta \lesssim 1$ meV (equivalent to an electron lifetime from picosecond to sub-nanosecond, see e.g., Refs. [39–41]). Meanwhile, $\omega$ should be on the order of MHz to GHz. Hence, the denominator in Eq. (B3) is dominantly determined by $E_{mn}$. One can see that the $(m, n)$ pair would make the major contribution to $\mathcal{C}$ when $E_{mn} = E_g$ with $E_g$ the bandgap. In this regard, we only consider this pair. Besides, we again use $[r_i]_{mn} \approx a_0$ and $\left\langle m \left| \frac{3r_i r_j - \delta_{ij} r^2}{r^5} \right| n \right\rangle \approx \frac{1}{a_0^3}$ with $a_0$ the Bohr radius. Then, one has

$$\mathcal{C} \approx \frac{g_S e^3 q}{2I(2I-1)} \frac{1}{4\pi\varepsilon_0 a_0^2} \frac{1}{E_g}, \tag{B4}$$

where $g_S = 2$ is the spin degeneracy of the electrons. We use this equation to estimate the strength of the NER/LQSE responses, and the results are listed in the fourth column of Table IV.

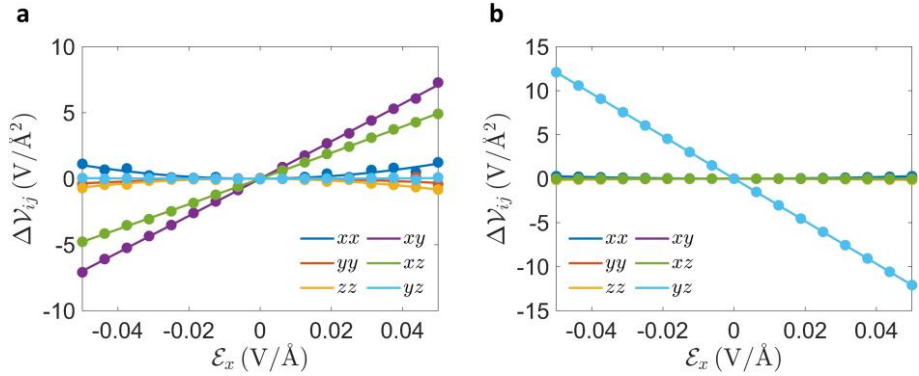

**Figure 6:** The EFG $\mathcal{V}$ tensor as a function of electric field $\mathcal{E}_x$ for (a) zinc-blende GaAs and (b) Sb defect in silicon.

**DFT Calculations.** Finally, the magnitude of the C tensor can be evaluated from DFT calculations.

To this purpose, we fit the $\mathcal{V}$-$\mathcal{E}$ curve from DFT calculations, and then one has [cf. Eq. (B1)]

$$\begin{aligned} \mathcal{C}_{ij}^p &= \left. \frac{\partial Q_{ij}}{\partial \mathcal{E}_p} \right|_{\mathcal{E}=0} \\ &= \left. \frac{eq}{2I(2I-1)} \frac{\partial \mathcal{V}_{ij}}{\partial \mathcal{E}_p} \right|_{\mathcal{E}=0} \end{aligned} \tag{B5}$$

The $\mathcal{C}$ tensor from DFT calculations are listed in the fifth column of Table IV (see also Figure S4).

One can see that for the NER/LQSE, which is the first-order response, the theoretical predictions from both the perturbation theory and the DFT calculations give the correct order of magnitude of the $\mathcal{C}$ tensor, as compared with experimental results.

**Table V:** The strength of ONQ response [in $\frac{2\pi \cdot \text{MHz}}{(\text{V/Å})^2}$] obtained from perturbation theory and DFT calculations.

|  | Perturbation Theory | DFT Calculation |
|---|---|---|
| $^{69}$Ga in wurtzite GaN | 6 | 1 |
| $^{75}$As in zinc blende GaAs | 24 | 20 |
| $^{123}$Sb defect in Si | 19 | 10 |

**2.3 The Second-Order Response**

Using several different systems as examples, we compared the $\mathcal{D}$ tensor estimated by second-order perturbation theory ($E_g - \omega_p = 0.2$ eV is used in Eq. (4) in the main text) and DFT calculations, and the results exhibit reasonable agreement (Table V). Considering that both the perturbation theory and the DFT calculation give the correct order of magnitude of the $\mathcal{C}$ tensor of the NER/LSQE (Table IV), we believe the theoretical predictions on the $\mathcal{D}$ tensor of the ONQ response should at least give the correct order of magnitude as well.

**Acknowledgments**. This work was supported by the Office of Naval Research MURI through grant #N00014-17-1-2661 and Honda Research Institute USA, Inc., through grant #031807-00001. A.R.B. acknowledges support from a National Science Foundation Graduate Research Fellowship under Grant No. DGE-174530. The calculations in this work were performed in part on the Texas Advanced Computing Center (TACC) and MIT Engaging cluster.

**Data Availability.** The authors declare that the main data supporting the findings of this study are available within the article and the appendix.

**Competing Interests.** The authors declare no competing interests.

**Author contributions.** H.X., P.C. and J.L. conceived the idea and designed the project. H.X. derived the theories and performed the ab initio calculations. C.L performed the master equation calculations. H.X. wrote the paper with contributions from all authors. P.C. and J.L. supervised the project. All authors analyzed the data and contributed to the discussions of the results.

# References

[1] M. Kjaergaard, M. E. Schwartz, J. Braumüller, P. Krantz, J. I. J. Wang, S. Gustavsson, and W. D. Oliver, *Superconducting Qubits: Current State of Play*, Https://Doi.Org/10.1146/Annurev-Conmatphys-031119-050605 **11**, 369 (2020).

[2] C. D. Bruzewicz, J. Chiaverini, R. McConnell, and J. M. Sage, *Trapped-Ion Quantum Computing: Progress and Challenges*, Appl. Phys. Rev. **6**, 021314 (2019).

[3] G. Wolfowicz, F. J. Heremans, C. P. Anderson, S. Kanai, H. Seo, A. Gali, G. Galli, and D. D. Awschalom, *Quantum Guidelines for Solid-State Spin Defects*, Nat. Rev. Mater. 2021 610 **6**, 906 (2021).

[4] A. Chatterjee, P. Stevenson, S. De Franceschi, A. Morello, N. P. de Leon, and F. Kuemmeth, *Semiconductor Qubits in Practice*, Nat. Rev. Phys. 2021 33 **3**, 157 (2021).

[5] M. Saffman, *Quantum Computing with Atomic Qubits and Rydberg Interactions: Progress and Challenges*, J. Phys. B At. Mol. Opt. Phys. **49**, 202001 (2016).

[6] S. Ebadi, A. Keesling, M. Cain, T. T. Wang, H. Levine, D. Bluvstein, G. Semeghini, A. Omran, J.-G. Liu, R. Samajdar, X.-Z. Luo, B. Nash, X. Gao, B. Barak, E. Farhi, S. Sachdev, N. Gemelke, L. Zhou, S. Choi, H. Pichler, S.-T. Wang, M. Greiner, V. Vuletic, and M. D. Lukin, *Quantum Optimization of Maximum Independent Set Using Rydberg Atom Arrays*, Science (80-. ). **376**, 1209 (2022).

[7] M. Steger, K. Saeedi, M. L. W. Thewalt, J. J. L. Morton, H. Riemann, N. V. Abrosimov, P. Becker, and H. J. Pohl, *Quantum Information Storage for over 180 s Using Donor Spins in a 28Si "Semiconductor Vacuum,"* Science (80-. ). **336**, 1280 (2012).

[8] K. Saeedi, S. Simmons, J. Z. Salvail, P. Dluhy, H. Riemann, N. V. Abrosimov, P. Becker, H. J. Pohl, J. J. L. Morton, and M. L. W. Thewalt, *Room-Temperature Quantum Bit Storage Exceeding 39 Minutes Using Ionized Donors in Silicon-28*, Science (80-. ). **342**, 830 (2013).

[9] F. Jelezko, T. Gaebel, I. Popa, M. Domhan, A. Gruber, and J. Wrachtrup, *Observation of Coherent Oscillation of a Single Nuclear Spin and Realization of a Two-Qubit Conditional Quantum Gate*, Phys. Rev. Lett. **93**, 130501 (2004).

[10] M. V. Gurudev Dutt, L. Childress, L. Jiang, E. Togan, J. Maze, F. Jelezko, A. S. Zibrov, P. R. Hemmer, and M. D. Lukin, *Quantum Register Based on Individual Electronic and Nuclear Spin Qubits in Diamond*, Science (80-. ). **316**, 1312 (2007).

[11] J. J. Pla, K. Y. Tan, J. P. Dehollain, W. H. Lim, J. J. L. L. Morton, F. A. Zwanenburg, D. N. Jamieson, A. S. Dzurak, and A. Morello, *High-Fidelity Readout and Control of a Nuclear Spin Qubit in Silicon*, Nature **496**, 334 (2013).

[12] J. J. Pla, F. A. Mohiyaddin, K. Y. Tan, J. P. Dehollain, R. Rahman, G. Klimeck, D. N. Jamieson, A. S. Dzurak, and A. Morello, *Coherent Control of a Single Si 29 Nuclear Spin Qubit*, Phys. Rev. Lett. **113**, 246801 (2014).

[13] G. D. Fuchs, G. Burkard, P. V. Klimov, and D. D. Awschalom, *A Quantum Memory Intrinsic to Single Nitrogen–Vacancy Centres in Diamond*, Nat. Phys. 2011 710 **7**, 789 (2011).

[14] P. London, J. Scheuer, J.-M. Cai, I. Schwarz, A. Retzker, M. B. Plenio, M. Katagiri, T. Teraji, S. Koizumi, J. Isoya, R. Fischer, L. P. Mcguinness, B. Naydenov, and F. Jelezko, *Detecting and Polarizing Nuclear Spins with Double Resonance on a Single Electron Spin*, Phys. Rev. Lett. **111**, 067601 (2013).

[15] A. J. Sigillito, A. M. Tyryshkin, T. Schenkel, A. A. Houck, and S. A. Lyon, *All-Electric Control of Donor Nuclear Spin Qubits in Silicon*, Nat. Nanotechnol. | **12**, (2017).

[16] S. Asaad, V. Mourik, B. Joecker, M. A. I. Johnson, A. D. Baczewski, H. R. Firgau, M. T. Mądzik, V. Schmitt, J. J. Pla, F. E. Hudson, K. M. Itoh, J. C. Mccallum, A. S. Dzurak, A. Laucht, and A. Morello, *Coherent Electrical Control of a Single High-Spin Nucleus in Silicon*, Nature **579**, (2020).

[17] R. Savytskyy, T. Botzem, I. F. de Fuentes, B. Joecker, F. E. Hudson, K. M. Itoh, A. M. Jakob, B. C. Johnson, D. N. Jamieson, A. S. Dzurak, and A. Morello, *An Electrically-Driven Single-Atom `flip-Flop' Qubit*, (2022).

[18]     J. Leng, F. Yang, and X.-B. Wang, *Nuclear Electric Resonance*, ArXiv Prepr. ArXiv2111.01424 (2021).

[19]     Y. Okazaki, I. Mahboob, K. Onomitsu, S. Sasaki, S. Nakamura, N.-H. H. Kaneko, and H. Yamaguchi, *Dynamical Coupling between a Nuclear Spin Ensemble and Electromechanical Phonons*, Nat. Commun. **9**, 1 (2018).

[20]     S. Maity, B. Pingault, G. Joe, M. Chalupnik, D. Assumpção, E. Cornell, L. Shao, and M. Lončar, *Mechanical Control of a Single Nuclear Spin*, Phys. Rev. X **12**, 11056 (2022).

[21]     M. L. Goldman, T. L. Patti, † D Levonian, S. F. Yelin, and M. D. Lukin, *Optical Control of a Single Nuclear Spin in the Solid State*, Phys. Rev. Lett. **124**, (2020).

[22]     Y. Puttisong, X. J. Wang, I. A. Buyanova, L. Geelhaar, H. Riechert, A. J. Ptak, C. W. Tu, and W. M. Chen, *Efficient Room-Temperature Nuclear Spin Hyperpolarization of a Defect Atom in a Semiconductor*, Nat. Commun. 2013 41 **4**, 1 (2013).

[23]     S. Yang, Y. Wang, D. D. B. B. Rao, T. H. Tran, A. S. Momenzadeh, M. Markham, D. J. Twitchen, P. Wang, W. Yang, R. Stöhr, P. Neumann, H. Kosaka, and J. Wrachtrup, *High-Fidelity Transfer and Storage of Photon States in a Single Nuclear Spin*, Nat. Photonics 2016 108 **10**, 507 (2016).

[24]     D. Serrano, J. Karlsson, A. Fossati, A. Ferrier, and P. Goldner, *All-Optical Control of Long-Lived Nuclear Spins in Rare-Earth Doped Nanoparticles*, Nat. Commun. 2018 91 **9**, 1 (2018).

[25]     O. Katz, R. Shaham, and O. Firstenberg, *Coupling Light to a Nuclear Spin Gas with a Two-Photon Linewidth of Five Millihertz*, Sci. Adv. **7**, eabe9164 (2021).

[26]     E. J. Sie, C. M. Nyby, C. D. Pemmaraju, S. J. Park, X. Shen, J. Yang, M. C. Hoffmann, B. K. Ofori-Okai, R. Li, A. H. Reid, S. Weathersby, E. Mannebach, N. Finney, D. Rhodes, D. Chenet, A. Antony, L. Balicas, J. Hone, T. P. Devereaux, T. F. Heinz, X. Wang, and A. M. Lindenberg, *An Ultrafast Symmetry Switch in a Weyl Semimetal*, Nature.

[27]     E. A. Mashkovich, K. A. Grishunin, R. M. Dubrovin, A. K. Zvezdin, R. V. Pisarev, and A. V. Kimel, *Terahertz-Light Driven Coupling of Antiferromagnetic Spins to Lattice*, Science (80-. ). **374**, 1608 (2021).

[28]     X. Li, T. Qiu, J. Zhang, E. Baldini, J. Lu, A. M. Rappe, and K. A. Nelson, *Terahertz Field–Induced Ferroelectricity in Quantum Paraelectric SrTiO3*, Science (80-. ). **364**, 1079 (2019).

[29]     J. W. McIver, B. Schulte, F. U. Stein, T. Matsuyama, G. Jotzu, G. Meier, and A. Cavalleri, *Light-Induced Anomalous Hall Effect in Graphene*, Nature Physics.

[30]     D. Layden, M. Chen, and P. Cappellaro, *Efficient Quantum Error Correction of Dephasing Induced by a Common Fluctuator*, Phys. Rev. Lett. **124**, 020504 (2020).

[31]     H. Xu, H. Wang, J. Zhou, and J. Li, *Pure Spin Photocurrent in Non-Centrosymmetric Crystals: Bulk Spin Photovoltaic Effect*, Nat. Commun. 2021 121 **12**, 1 (2021).

[32]     H. Xu, H. Wang, J. Zhou, Y. Guo, J. Kong, and J. Li, *Colossal Switchable Photocurrents in Topological Janus Transition Metal Dichalcogenides*, Npj Comput. Mater. **7**, 1 (2021).

[33]     P. Hohenberg and W. Kohn, *Inhomogeneous Electron Gas*, Phys. Rev. **136**, B864 (1964).

[34]     W. Kohn and L. J. Sham, *Self-Consistent Equations Including Exchange and Correlation Effects*, Phys. Rev. **140**, A1133 (1965).

[35]     F. Helmchen and W. Denk, *Deep Tissue Two-Photon Microscopy*, Nat. Methods 2005 212 **2**, 932 (2005).

[36]     G. Kurizki, P. Bertet, Y. Kubo, K. Mølmer, D. Petrosyan, P. Rabl, and J. Schmiedmayer, *Quantum Technologies with Hybrid Systems*, Proc. Natl. Acad. Sci. U. S. A. **112**, 3866 (2015).

[37]     H. M. Petrilli, P. E. Blöchl, P. Blaha, and K. Schwarz, *Electric-Field-Gradient Calculations Using the Projector Augmented Wave Method*, Phys. Rev. B **57**, 14690 (1998).

[38]     J. Y. Tsao, S. Chowdhury, M. A. Hollis, D. Jena, N. M. Johnson, K. A. Jones, R. J. Kaplar, S. Rajan, C. G. Van de Walle, E. Bellotti, C. L. Chua, R. Collazo, M. E. Coltrin, J. A. Cooper, K. R. Evans, S. Graham, T. A. Grotjohn, E. R. Heller, M. Higashiwaki, M. S. Islam, P. W. Juodawlkis, M. A. Khan, A. D. Koehler, J. H. Leach, U. K. Mishra, R. J. Nemanich, R. C. N. Pilawa-Podgurski, J. B. Shealy, Z. Sitar, M. J. Tadjer, A. F. Witulski, M. Wraback, and J. A. Simmons, *Ultrawide-Bandgap Semiconductors: Research Opportunities and Challenges*, Adv. Electron. Mater. **4**, 1600501 (2018).

[39] S. Gupta, M. Y. Frankel, J. A. Valdmanis, J. F. Whitaker, G. A. Mourou, F. W. Smith, and A. R. Calawa, *Subpicosecond Carrier Lifetime in GaAs Grown by Molecular Beam Epitaxy at Low Temperatures*, Appl. Phys. Lett. **59**, 3276 (1991).

[40] Z. Z. Bandić, P. M. Bridger, E. C. Piquette, and T. C. McGill, *Electron Diffusion Length and Lifetime in P-Type GaN*, Appl. Phys. Lett. **73**, 3276 (1998).

[41] N. H. Protik and B. Kozinsky, *Electron-Phonon Drag Enhancement of Transport Properties from a Fully Coupled Ab Initio Boltzmann Formalism*, Phys. Rev. B **102**, 245202 (2020).

[42] P. C. Maurer, G. Kucsko, C. Latta, L. Jiang, N. Y. Yao, S. D. Bennett, F. Pastawski, D. Hunger, N. Chisholm, M. Markham, others, D. J. Twitchen, J. I. Cirac, and M. D. Lukin, *Room-Temperature Quantum Bit Memory Exceeding One Second*, Science (80-. ). **336**, 1283 (2012).

[43] P. Neumann, J. Beck, M. Steiner, F. Rempp, H. Fedder, P. R. Hemmer, J. Wrachtrup, and F. Jelezko, *Single-Shot Readout of a Single Nuclear Spin*, Science (80-. ). **329**, 542 (2010).

[44] L. A. Lyon, C. D. Keating, A. P. Fox, B. E. Baker, L. He, S. R. Nicewarner, S. P. Mulvaney, and M. J. Natan, *Raman Spectroscopy*, Anal. Chem. **70**, (1998).

[45] X. Cong, X. L. Liu, M. L. Lin, and P. H. Tan, *Application of Raman Spectroscopy to Probe Fundamental Properties of Two-Dimensional Materials*, Npj 2D Mater. Appl. 2020 41 **4**, 1 (2020).

[46] W. H. Renninger, P. Kharel, R. O. Behunin, and P. T. Rakich, *Bulk Crystalline Optomechanics*, Nat. Phys. 2018 146 **14**, 601 (2018).

[47] J. M. Taylor, C. M. Marcus, and M. D. Lukin, *Long-Lived Memory for Mesoscopic Quantum Bits*, Phys. Rev. Lett. **90**, 4 (2003).

[48] D. A. Gangloff, G. Éthier-Majcher, C. Lang, E. V. Denning, J. H. Bodey, D. M. Jackson, E. Clarke, M. Hugues, C. Le Gall, and M. Atatüre, *Quantum Interface of an Electron and a Nuclear Ensemble*, Science (80-. ). **364**, 62 (2019).

[49] D. M. Jackson, D. A. Gangloff, J. H. Bodey, L. Zaporski, C. Bachorz, E. Clarke, M. Hugues, C. Le Gall, and M. Atatüre, *Quantum Sensing of a Coherent Single Spin Excitation in a Nuclear Ensemble*, Nat. Phys. 2021 175 **17**, 585 (2021).

[50] D. A. Gangloff, L. Zaporski, J. H. Bodey, C. Bachorz, D. M. Jackson, G. Éthier-Majcher, C. Lang, E. Clarke, M. Hugues, C. Le Gall, and M. Atatüre, *Witnessing Quantum Correlations in a Nuclear Ensemble via an Electron Spin Qubit*, Nat. Phys. 2021 1711 **17**, 1247 (2021).

[51] J. Choi, H. Zhou, H. S. Knowles, R. Landig, S. Choi, and M. D. Lukin, *Robust Dynamic Hamiltonian Engineering of Many-Body Spin Systems*, Phys. Rev. X **10**, 31002 (2020).

[52] H. Zhou, J. Choi, S. Choi, R. Landig, A. M. Douglas, J. Isoya, F. Jelezko, S. Onoda, H. Sumiya, P. Cappellaro, others, H. S. Knowles, H. Park, and M. D. Lukin, *Quantum Metrology with Strongly Interacting Spin Systems*, Phys. Rev. X **10**, 31003 (2020).

[53] H. Xu, *To Be Published*, (n.d.).

[54] S. Probst, A. Tkalčec, H. Rotzinger, D. Rieger, J. M. Le Floch, M. Goryachev, M. E. Tobar, A. V. Ustinov, and P. A. Bushev, *Three-Dimensional Cavity Quantum Electrodynamics with a Rare-Earth Spin Ensemble*, Phys. Rev. B - Condens. Matter Mater. Phys. **90**, 100404 (2014).

[55] H. Goto, K. Ichimura, and S. Nakamura, *Experimental Determination of Intracavity Losses of Monolithic Fabry-Perot Cavities Made of $Pr^{3+}:Y_2SiO_5$*, Opt. Express, Vol. 18, Issue 23, Pp. 23763-23775 **18**, 23763 (2010).

[56] V. Huet, A. Rasoloniaina, P. Guillemé, P. Rochard, P. Féron, M. Mortier, A. Levenson, K. Bencheikh, A. Yacomotti, and Y. Dumeige, *Millisecond Photon Lifetime in a Slow-Light Microcavity*, Phys. Rev. Lett. **116**, 133902 (2016).

[57] I. Esmaeil Zadeh, J. W. N. N. Los, R. B. M. M. Gourgues, V. Steinmetz, G. Bulgarini, S. M. Dobrovolskiy, V. Zwiller, and S. N. Dorenbos, *Single-Photon Detectors Combining High Efficiency, High Detection Rates, and Ultra-High Timing Resolution*, Apl Photonics **2**, 111301 (2017).

[58] V. B. Verma, B. Korzh, A. B. Walter, A. E. Lita, R. M. Briggs, M. Colangelo, Y. Zhai, E. E. Wollman, A. D. Beyer, J. P. Allmaras, H. Vora, D. Zhu, E. Schmidt, A. G. Kozorezov, K. K. Berggren, R. P. Mirin, S. W. Nam, M. D. Shaw, and others, *Single-Photon Detection in the Mid-Infrared up to 10 $µ$ m*

[59] P. Arrangoiz-Arriola, E. A. Wollack, Z. Wang, M. Pechal, W. Jiang, T. P. McKenna, J. D. Witmer, R. Van Laer, and A. H. Safavi-Naeini, *Resolving the Energy Levels of a Nanomechanical Oscillator*, Nature **571**, 537 (2019).

[60] A. Meldrum, P. Bianucci, and F. Marsiglio, *Modification of Ensemble Emission Rates and Luminescence Spectra for Inhomogeneously Broadened Distributions of Quantum Dots Coupled to Optical Microcavities*, Opt. Express **18**, 10230 (2010).

[61] S. Probst, H. Rotzinger, S. Wünsch, P. Jung, M. Jerger, M. Siegel, A. V. Ustinov, and P. A. Bushev, *Anisotropic Rare-Earth Spin Ensemble Strongly Coupled to a Superconducting Resonator*, Phys. Rev. Lett. **110**, 157001 (2013).

[62] Y. Reshitnyk, M. Jerger, and A. Fedorov, *3D Microwave Cavity with Magnetic Flux Control and Enhanced Quality Factor*, EPJ Quantum Technol. **3**, 1 (2016).

[63] L. A. Williamson, Y.-H. Chen, and J. J. Longdell, *Magneto-Optic Modulator with Unit Quantum Efficiency*, Phys. Rev. Lett. **113**, 203601 (2014).

[64] J. P. Covey, A. Sipahigil, and M. Saffman, *Microwave-to-Optical Conversion via Four-Wave Mixing in a Cold Ytterbium Ensemble*, Phys. Rev. A **100**, 012307 (2019).

[65] J. R. Everts, G. G. G. King, N. J. Lambert, S. Kocsis, S. Rogge, and J. J. Longell, *Ultrastrong Coupling between a Microwave Resonator and Antiferromagnetic Resonances of Rare-Earth Ion Spins*, Phys. Rev. B **101**, 214414 (2020).

[66] J. R. Johansson, P. D. Nation, and F. Nori, *QuTiP: An Open-Source Python Framework for the Dynamics of Open Quantum Systems*, Comput. Phys. Commun. **183**, 1760 (2012).

[67] J. R. Johansson, P. D. Nation, and F. Nori, *QuTiP 2: A Python Framework for the Dynamics of Open Quantum Systems*, Comput. Phys. Commun. **184**, 1234 (2013).

[68] C. Li and P. Cappellaro, *Telecom Photon Interface of Solid-State Quantum Nodes*, J. Phys. Commun. **3**, 95016 (2019).

[69] C. O'Brien, N. Lauk, S. Blum, G. Morigi, and M. Fleischhauer, *Interfacing Superconducting Qubits and Telecom Photons via a Rare-Earth-Doped Crystal*, Phys. Rev. Lett. **113**, 063603 (2014).

[70] K. Svensson, J. Lundqvist, E. Campos, al -, S. Forouzanfar, R. Mansour, E. Abdel-Rahman, N. Lauk, N. Sinclair, S. Barzanjeh, J. P. Covey, M. Saffman, M. Spiropulu, and C. Simon, *Perspectives on Quantum Transduction*, Quantum Sci. Technol. **5**, 020501 (2020).

[71] J. Zhang, K. Peng, and S. L. Braunstein, *Quantum-State Transfer from Light to Macroscopic Oscillators*, Phys. Rev. A **68**, 013808 (2003).

[72] J. Bochmann, A. Vainsencher, D. D. Awschalom, and A. N. Cleland, *Nanomechanical Coupling between Microwave and Optical Photons*, Nat. Phys. 2013 911 **9**, 712 (2013).

[73] R. W. Andrews, R. W. Peterson, T. P. Purdy, K. Cicak, R. W. Simmonds, C. A. Regal, and K. W. Lehnert, *Bidirectional and Efficient Conversion between Microwave and Optical Light*, Nat. Phys. 2014 104 **10**, 321 (2014).

[74] T. Bagci, A. Simonsen, S. Schmid, L. G. Villanueva, E. Zeuthen, J. Appel, J. M. Taylor, A. Sørensen, K. Usami, A. Schliesser, and E. S. Polzik, *Optical Detection of Radio Waves through a Nanomechanical Transducer*, Nat. 2014 5077490 **507**, 81 (2014).

[75] J. Verdú, H. Zoubi, C. Koller, J. Majer, H. Ritsch, and J. Schmiedmayer, *Strong Magnetic Coupling of an Ultracold Gas to a Superconducting Waveguide Cavity*, Phys. Rev. Lett. **103**, 043603 (2009).

[76] A. Imamoğlu, *Cavity QED Based on Collective Magnetic Dipole Coupling: Spin Ensembles as Hybrid Two-Level Systems*, Phys. Rev. Lett. **102**, 083602 (2009).

[77] M. Hafezi, Z. Kim, S. L. Rolston, L. A. Orozco, B. L. Lev, and J. M. Taylor, *Atomic Interface between Microwave and Optical Photons*, Phys. Rev. A - At. Mol. Opt. Phys. **85**, 020302 (2012).

[78] J. Han, T. Vogt, C. Gross, D. Jaksch, M. Kiffner, and W. Li, *Coherent Microwave-to-Optical Conversion via Six-Wave Mixing in Rydberg Atoms*, Phys. Rev. Lett. **120**, 093201 (2018).

[79] T. Vogt, C. Gross, J. Han, S. B. Pal, M. Lam, M. Kiffner, and W. Li, *Efficient Microwave-to-Optical Conversion Using Rydberg Atoms*, Phys. Rev. A **99**, 023832 (2019).

[80] J. G. Bartholomew, J. Rochman, T. Xie, J. M. Kindem, A. Ruskuc, I. Craiciu, M. Lei, and A. Faraon, *On-Chip Coherent Microwave-to-Optical Transduction Mediated by Ytterbium in YVO 4*, (2006).

[81] R. Hisatomi, A. Osada, Y. Tabuchi, T. Ishikawa, A. Noguchi, R. Yamazaki, K. Usami, and Y. Nakamura, *Bidirectional Conversion between Microwave and Light via Ferromagnetic Magnons*, Phys. Rev. B **93**, 174427 (2016).

[82] J. R. Everts, M. C. Berrington, R. L. Ahlefeldt, and J. J. Longdell, *Microwave to Optical Photon Conversion via Fully Concentrated Rare-Earth-Ion Crystals*, Phys. Rev. A **99**, 063830 (2019).

[83] P. A. Rinck, *Magnetic Resonance in Medicine : A Critical Introduction : The Basic Text Book of the European Magnetic Resonance Forum*, 375 (2018).

[84] S. C. Bushong and G. D. Clarke, *Magnetic Resonance Imaging : Physical and Biological Principles*, 513 (n.d.).

[85] M. Ono, J. Ishihara, G. Sato, Y. Ohno, and H. Ohno, *Coherent Manipulation of Nuclear Spins in Semiconductors with an Electric Field*, Appl. Phys. Express **6**, 33002 (2013).

[86] R. W. P. P. Drever, J. L. Hall, F. V. Kowalski, J. J. Hough, G. M. Ford, A. J. Munley, and H. Ward, *Laser Phase and Frequency Stabilization Using an Optical Resonator*, Appl. Phys. B **31**, 97 (1983).

[87] J. L. Hall, *Stabilizing Lasers for Applications in Quantum Optics*, in *Quantum Optics IV* (Springer, 1986), pp. 273–284.

[88] C. Salomon, D. Hils, and J. L. Hall, *Laser Stabilization at the Millihertz Level*, JOSA B, Vol. 5, Issue 8, Pp. 1576-1587 **5**, 1576 (1988).

[89] T. Day, E. K. Gustafson, and R. L. Byer, *Sub-Hertz Relative Frequency Stabilization of Two-Diode Laser-Pumped Nd: YAG Lasers Locked to a Fabry-Perot Interferometer*, IEEE J. Quantum Electron. **28**, 1106 (1992).

[90] B. C. Young, F. C. Cruz, W. M. Itano, and J. C. Bergquist, *Visible Lasers with Subhertz Linewidths*, Phys. Rev. Lett. **82**, 3799 (1999).

[91] J. Burris and T. J. McIlrath, *Theoretical Study Relating the Two-Photon Absorption Cross Section to the Susceptibility Controlling Four-Wave Mixing*, JOSA B, Vol. 2, Issue 8, Pp. 1313-1317 **2**, 1313 (1985).

[92] M. M. Choy and R. L. Byer, *Accurate Second-Order Susceptibility Measurements of Visible and Infrared Nonlinear Crystals*, Phys. Rev. B **14**, 1693 (1976).

[93] Y.-R. Shen, *Principles of Nonlinear Optics*, (1984).

[94] G. P. Agrawal and R. W. Boyd, *Contemporary Nonlinear Optics*, Contemp. Nonlinear Opt. (1992).

[95] *Third-Order Nonlinear Optical Coefficients of Si and GaAs in the Near-Infrared Spectral Region | IEEE Conference Publication | IEEE Xplore*, https://ieeexplore.ieee.org/document/8427144.

[96] M. S. Dresselhaus, *Solid State Physics Part II Optical Properties of Solids*, (n.d.).

[97] L. V Keldysh, *IONIZATION IN THE FIELD OF A STRONG ELECTROMAGNETIC WAVE*, J. Exptl. Theor. Phys. **20**, 1945 (1965).

[98] C. B. Schaffer, A. Brodeur, and E. Mazur, *Laser-Induced Breakdown and Damage in Bulk Transparent Materials Induced by Tightly Focused Femtosecond Laser Pulses*, Meas. Sci. Technol. **12**, 1784 (2001).

[99] I. S. Grudinin, A. B. Matsko, A. A. Savchenkov, D. Strekalov, V. S. Ilchenko, and L. Maleki, *Ultra High Q Crystalline Microcavities*, Opt. Commun. **265**, 33 (2006).

[100] H. Choi, M. Heuck, and D. Englund, *Self-Similar Nanocavity Design with Ultrasmall Mode Volume for Single-Photon Nonlinearities*, Phys. Rev. Lett. **118**, 223605 (2017).

[101] D. I. Schuster, A. A. Houck, J. A. Schreier, A. Wallraff, J. M. Gambetta, A. Blais, L. Frunzio, J. Majer, B. Johnson, M. H. Devoret, S. M. Girvin, and R. J. Schoelkopf, *Resolving Photon Number States in a Superconducting Circuit*, Nat. 2006 4457127 **445**, 515 (2007).

[102] J. J. Viennot, X. Ma, and K. W. Lehnert, *Phonon-Number-Sensitive Electromechanics*, Phys. Rev. Lett. **121**, 183601 (2018).

[103] L. R. Sletten, B. A. Moores, J. J. Viennot, and K. W. Lehnert, *Resolving Phonon Fock States in a Multimode Cavity with a Double-Slit Qubit*, Phys. Rev. X **9**, 21056 (2019).

[104] I. Schuster, A. Kubanek, A. Fuhrmanek, T. Puppe, P. W. H. Pinkse, K. Murr, and G. Rempe, *Nonlinear Spectroscopy of Photons Bound to One Atom*, Nat. Phys. 2008 45 **4**, 382 (2008).

[105] A. Kubanek, A. Ourjoumtsev, I. Schuster, M. Koch, P. W. H. Pinkse, K. Murr, and G. Rempe, *Two-Photon Gateway in One-Atom Cavity Quantum Electrodynamics*, Phys. Rev. Lett. **101**, 203602 (2008).

[106] L. Z. Tan, F. Zheng, S. M. Young, F. Wang, S. Liu, and A. M. Rappe, *Shift Current Bulk Photovoltaic Effect in Polar Materials-Hybrid and Oxide Perovskites and Beyond*, Npj Comput. Mater. **2**, 1 (2016).

[107] S. Y. Yang, J. Seidel, S. J. Byrnes, P. Shafer, C.-H. H. Yang, M. D. Rossell, P. Yu, Y.-H. H. Chu, J. F. Scott, J. W. Ager, L. W. Martin, R. Ramesh, and others, *Above-Bandgap Voltages from Ferroelectric Photovoltaic Devices*, Nat. Nanotechnol. **5**, 143 (2010).

[108] D. Meiser, J. Ye, D. R. Carlson, and M. J. Holland, *Prospects for a Millihertz-Linewidth Laser*, Phys. Rev. Lett. **102**, 163601 (2009).

[109] J. E. Sipe and E. Ghahramani, *Nonlinear Optical Response of Semiconductors in the Independent-Particle Approximation*, Phys. Rev. B **48**, 11705 (1993).

[110] H. Xu, H. Wang, and J. Li, *Abnormal Nonlinear Optical Responses on the Surface of Topological Materials*, Npj Comput. Mater. **8**, 1 (2022).

[111] R. C. Eckardt, *Nonlinear Optics, Basics| $χ$ (2)--Harmonic Generation*, (2005).

[112] P. Giannozzi, S. Baroni, N. Bonini, M. Calandra, R. Car, C. Cavazzoni, D. Ceresoli, G. L. Chiarotti, M. Cococcioni, I. Dabo, A. Dal Corso, S. De Gironcoli, S. Fabris, G. Fratesi, R. Gebauer, U. Gerstmann, C. Gougoussis, A. Kokalj, M. Lazzeri, L. Martin-Samos, N. Marzari, F. Mauri, R. Mazzarello, S. Paolini, A. Pasquarello, L. Paulatto, C. Sbraccia, S. Scandolo, G. Sclauzero, A. P. Seitsonen, A. Smogunov, P. Umari, and R. M. Wentzcovitch, *QUANTUM ESPRESSO: A Modular and Open-Source Software Project for Quantum Simulations of Materials Related Content QUANTUM ESPRESSO: A Modular and Open-Source Software Project for Quantum Simulations of Materials*, J. Phys. Condens. Matter **21**, 395502 (2009).

[113] P. Giannozzi, O. Andreussi, T. Brumme, O. Bunau, M. Buongiorno Nardelli, M. Calandra, R. Car, C. Cavazzoni, D. Ceresoli, M. Cococcioni, N. Colonna, I. Carnimeo, A. Dal Corso, S. De Gironcoli, P. Delugas, R. A. Distasio, A. Ferretti, A. Floris, G. Fratesi, G. Fugallo, R. Gebauer, U. Gerstmann, F. Giustino, T. Gorni, J. Jia, M. Kawamura, H. Y. Ko, A. Kokalj, E. Kücükbenli, M. Lazzeri, M. Marsili, N. Marzari, F. Mauri, N. L. Nguyen, H. V. Nguyen, A. Otero-De-La-Roza, L. Paulatto, S. Poncé, D. Rocca, R. Sabatini, B. Santra, M. Schlipf, A. P. Seitsonen, A. Smogunov, I. Timrov, T. Thonhauser, P. Umari, N. Vast, X. Wu, and S. Baroni, *Advanced Capabilities for Materials Modelling with Quantum ESPRESSO*, J. Phys. Condens. Matter **29**, 465901 (2017).

[114] J. P. Perdew, K. Burke, and M. Ernzerhof, *Generalized Gradient Approximation Made Simple*, Phys. Rev. Lett. **77**, 3865 (1996).

[115] P. E. Blöchl, *Projector Augmented-Wave Method*, Phys. Rev. B **50**, 17953 (1994).

[116] R. D. King-Smith and D. Vanderbilt, *Theory of Polarization of Crystalline Solids*, Phys. Rev. B **47**, 1651 (1993).

[117] R. Resta and D. Vanderbilt, *Theory of Polarization: A Modern Approach*, Top. Appl. Phys. **105**, 31 (2007).

[118] T. Charpentier, *The PAW/GIPAW Approach for Computing NMR Parameters: A New Dimension Added to NMR Study of Solids*, Solid State Nucl. Magn. Reson. **40**, 1 (2011).

[119] V. Ivády, I. A. Abrikosov, and A. Gali, *First Principles Calculation of Spin-Related Quantities for Point Defect Qubit Research*, Npj Comput. Mater. **4**, 1 (2018).

[120] Á. Gali, *Ab Initio Theory of the Nitrogen-Vacancy Center in Diamond*, Nanophotonics **8**, 1907 (2019).

[121] M. Pfender, N. Aslam, P. Simon, D. Antonov, G. Thiering, S. Burk, F. de Oliveira, A. Denisenko, H. Fedder, J. Meijer, and others, *Protecting a Diamond Quantum Memory by Charge State Control*, Nano Lett. **17**, 5931 (2017).

[122] G. Wang, A. R. Barr, H. Tang, M. Chen, C. Li, H. Xu, J. Li, and P. Cappellaro, *Characterizing Temperature and Strain Variations with Qubit Ensembles for Their Robust Coherence Protection*, ArXiv Prepr. ArXiv2205.02790 (2022).

[123] H. Tang, A. R. Barr, G. Wang, P. Cappellaro, and J. Li, *First-Principles Calculation of the Temperature-Dependent Transition Energies in Spin Defects*, ArXiv Prepr. ArXiv2205.02791 (2022).

[124] D. Gill and N. Bloembergen, *Linear Stark Splitting of Nuclear Spin Levels in GaAs*, Phys. Rev. **129**, 2398 (1963).

[125] R. W. Dixon and N. Bloembergen, *Electrically Induced Perturbations of Halogen Nuclear Quadrupole Interactions in Polycrystalline Compounds. I. Phenomenological Theory and Experimental Results*, J. Chem. Phys. **41**, 1720 (1964).

[126] H. Xu, H. Wang, J. Zhou, and J. Li, *Pure Spin Photocurrent in Non-Centrosymmetric Crystals: Bulk Spin Photovoltaic Effect*, (2020).